%% file: Strutz_Curtis_GeophysicalBED.tex
\documentclass{article}

\usepackage{arxiv}
\usepackage{natbib}

\input{utils.tex}

\title{Variational Bayesian experimental design for geophysical applications}
\author{Dominik Strutz$^1$ and Andrew Curtis $^1$ \\
  $^1$ School of GeoSciences, University of Edinburgh, Edinburgh EH9 3FE, UK. E-mail: dstrutz@ed.ac.uk
  }
\date{June 2023}

\renewcommand{\undertitle}{A Preprint submitted to Geophysical Journal International in June 2023}

\hypersetup{
  pdfauthor={Dominik Strutz and Andrew Curtis},
  pdftitle={Variational Bayesian experimental design for geophysical applications},
}

\begin{document}

\maketitle

\begin{abstract}
    \input{./abstract.tex}
\end{abstract}


\section{Introduction}\label{sec:introduction}
    \input{./introduction.tex}

\section{Bayesian Experimental Design}\label{sec:BED}
\input{./BED.tex}

\section{EIG Estimation}\label{sec:EIG_est}
\input{./EIG_estimation.tex}

\section{EIG Optimisation}\label{sec:EIG_opt}
\input{./EIG_optimisation.tex}

\section{Advanced Concepts}\label{sec:advanced_concepts}
\input{./advanced_concepts.tex}

\section{Geophysical Applications of Bayesian Optimal Design Methods}\label{sec:previous_work}
\input{./previous_work.tex}

\section{Applications}\label{sec:applications}
\input{./applications_srcloc.tex}
\input{./applications_avo.tex}

\section{Discussion}\label{sec:discussion}
\input{./discussion.tex}

\section{Conclusions}\label{sec:conclusions}
\input{./conclusions.tex}

\section{Data Availability}\label{sec:data_availability}
\input{./data_availability.tex}

\section{Acknowledgements}\label{sec:acknowledgements}
\input{./acknowledgements.tex}

\bibliographystyle{gji}
\bibliography{bibliography.bib}

\end{document}

%% file: utils.tex
\usepackage{amsmath}
\usepackage{amssymb}

\usepackage{hyperref}

\usepackage{graphicx} 
\graphicspath{{./}}
\usepackage{subcaption}

\usepackage{xspace} 

\binoppenalty=\maxdimen
\relpenalty=\maxdimen

\newcommand{\eg}{e.\,g.,\,}

\newcommand{\mitbf}[1]{  
  \hbox{\mathversion{bold}$#1$}}

\newcommand{\com}[1]{\ignorespaces}
\newcommand{\mo}[1]{\operatorname{{#1}}}

\renewcommand{\d}{$\mitbf{d}$\xspace} 
\newcommand{\m}{$\mitbf{m}$\xspace}
\newcommand{\F}{$F$\xspace}
\newcommand{\design}{$\xi$\xspace}

\newcommand{\posterior}{$p(\mitbf{m} \, | \, \mitbf{d}, \xi)$\xspace}
\newcommand{\evidence}{$p(\mitbf{d} \, | \, \xi)$\xspace}
\newcommand{\likelihood}{$p(\mitbf{d} \, | \, \mitbf{m}, \xi)$\xspace}
\newcommand{\prior}{$p(\mitbf{m})$\xspace}

\newcommand{\IG}{$\mo{IG}$\xspace}
\newcommand{\EIG}{$\mo{EIG}$\xspace}

\newcommand\tento[1]{\ensuremath{{\times}10^{#1}}}

%% file: abstract.tex
In geophysical experiments or surveys, recorded data are used to constrain target properties or dynamics of the planetary subsurface, oceans, cryosphere or atmosphere. The exact choice of experimental design controls how much and precisely what information is transferred to target variables. Bayesian experimental design methods quantify, characterise, and maximise target information. This constitutes a macro-optimisation problem in which we optimise the design of the set of Bayesian inference problems that we might encounter post-experiment. Typical design parameters that can be varied are source and sensor types and locations, and the choice of modelling or data processing methods to be applied to the data; these may all be optimised subject to various cost constraints. 
This paper introduces variational design methods that are novel to Geophysics, and discusses their benefits and limitations in the context of geophysical applications and more established design methods. Variational methods rely on functional approximations to probability distributions and model-data relationships, and have recently come to prominence due to their importance in machine learning applications. They can be used to design experiments that best resolve either all model parameters, or the answer to specific questions about the system to be interrogated. The methods are tested in three schematic geophysical applications: (i) estimating a source location given arrival times at sensor locations, and (ii) estimating the contrast in seismic wavefield velocity across a stratal interface given measurements of the amplitudes of seismic wavefield reflections from that interface, and (iii) designing a survey to best constrain CO2 saturation in a subsurface storage scenario. Variational methods allow the value of an experiment to be calculated and optimised simultaneously, which results in substantial savings in computational cost. In the context of designing a survey to best constrain CO2 in a subsurface storage scenario, we show that optimal designs may change substantially depending on the questions of interest. Overall, this work indicates that optimal design methods should be used more widely in Geophysics, as they are in other scientifically advanced fields.

%% file: introduction.tex
Every geophysical investigation that collects data is an experiment, usually intended to estimate parameters that describe the properties of natural systems. How the experimental data are collected significantly influences which parameters can be resolved and how much confidence can be placed in the results. It is well known that the expected results can be improved by explicitly optimising the experimental design.

The field of optimal experimental design (OED) has a long history in industrial processes (Taguchi methods, \citet{Kiefer1959-rs, Atkinson1975-ts}) and had its first geophysical application in 1977, optimising receiver placement for locating seismic sources \citep{Kijko1977-ph, Kijko1977-zj}. The field has developed significantly in many areas since then: in Geophysics, OED has been used to design more sophisticated source location experiments ( \cite{Rabinowitz1990-cw,Rabinowitz2000-su,Steinberg1995-ox,Curtis2004-uq, Bloem2020-gp,Toledo2020-xn, Rawlinson2012-cs}), seismic tomography surveys (\citet{Curtis1997-iq, Curtis1999-hf, Curtis1999-mg,Liner1999-lv,Gibson2002-zr,Curtis2004-uq,Brenders2007-wv,Haber2008-ms,Ajo-Franklin2009-xx, Coles2009-rh, Maurer2009-cg,Khodja2010-st, Coles2011-dh,Djikpesse2012-mc,Coles2013-vw,Bernauer2014-gg,Maurer2017-tg,Nuber2017-dj, Krampe2021-sx}), reflected wave amplitude inversions (\citet{Guest2011-eu, Guest2010-zu, Guest2009-gi,Van_Den_Berg2003-bn,Van_Den_Berg2005-lq}), electromagnetic and electrical resistivity tomography (\citet{Maurer1998-vd, Maurer2000-zx, Stummer2002-ty, Furman2004-od, Stummer2004-lm, Wilkinson2006-kh, Oldenborger2009-rd, Maurer2010-jt, Wilkinson2012-fg, Qiang2022-ow, Coles2009-rh}), electrical impedance tomography (\citet{Hyvonen2014-pi}), expert elicitation (\citet{Curtis2004-wb, Runge2013-wb}), contaminant transport (\citet{Alexanderian2014-rs, Alexanderian2018-se, Zhang2015-sp}), $CO_2$ monitoring \citep{Romdhane2018-or}, array design \citep{Muir2021-pk}, and many others.

\begin{figure*}
    \includegraphics[width=\textwidth]{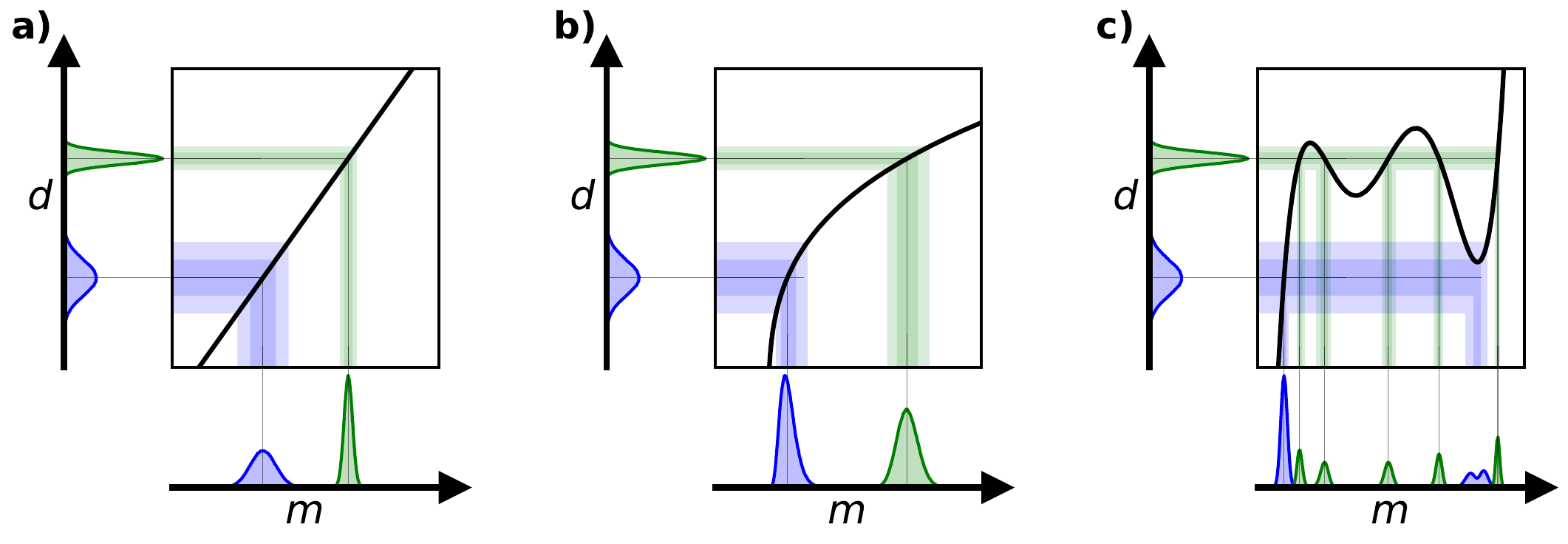}
    \caption{
        The green and blue Gaussian distributions on the $d$-axis represent two measurements with different uncertainties. Distributions on the $m$-axis represent the resulting back-projections of the two measurements through (a) linear, (b) slightly non-linear, and (c) more strongly non-linear model-data relationships (black lines).
    }
    \label{fig:oed-introduction}
\end{figure*}

There are three prerequisites for any experimental design problem: first, a function (physical or empirical) that relates the vector of model parameters \m to the vector of synthetic observations \d, called the \textit{forward function}. This relationship 
\begin{align}
\mitbf{d} = F(\mitbf{m}) \label{eqn:general_relation}
\end{align}
can be linear, but in most geophysical problems, $F$ is nonlinear. Second, we require a description of what is already known about the values of all parameters in vector \m, which are necessary to evaluate the forward function; this prior knowledge is usually described by a probability distribution function (pdf -- a probability density if variables are continuous rather than discrete), called the \textit{prior pdf}. Third, a pdf describing the probability of observing a datum \d if any particular set of values for a model parameter vector \m were true. The latter is commonly referred to as the \textit{likelihood}, and is often approximated by a Gaussian with mean $F(\mitbf{m})$ and variance corresponding to the measurement uncertainty. Using Bayesian inference, these three states of knowledge can be combined with information about \m derived from any future measured data set, the resultant state of information being called the \textit{posterior pdf}.

Both the content of data vector \d and the form of \F are significantly influenced by how an experiment is set up, in other words, by the experimental design. The appropriate approach to design an experiment depends on the nature of the forward function $F(\mitbf{m})$. Fig. \ref{fig:oed-introduction}(a) shows an example of a linear model-data relation. The observed data (green and blue) and their uncertainties (here, a Gaussian distribution) on the vertical axis are mapped into the model parameter domain by back-projecting the data uncertainty through \F onto the parameter axis $m$. Clearly, a more certain observation (green) yields a more certain model parameter estimate (a narrower pdf on the horizontal axis)  than a less certain observation (blue). However, the data uncertainty ranges are also scaled by the reciprocal slope of the forward function. Accordingly, optimal experimental design for linear forward functions varies the design so as to maximise the gradient (or higher dimensional equivalents) since a higher slope results in more accurate model parameter estimates given the same data uncertainties. The constant gradient of a linear mapping also implies that the optimal design is independent of the true parameter values or measured data, since if the gradient is maximised for one parameter value, it is maximised for all values. This property is important because it implies that we do not need a good estimate of the data, or of the Earth's properties prior to optimising the design.

The aforementioned characteristics gradually deteriorate as \F deviates from a linear function. Most properties still hold in a related sense for slightly non-linear functions, as shown in Fig. \ref{fig:oed-introduction}(b). Each measured datum still maps to a single point in model parameter space, and provided data uncertainties are small, the slope of the function at the back-projection point is usually the most influential factor affecting how measurement uncertainties relate to model uncertainties. However, unlike for linear functions, the slope now depends on the model parameter values, so parameter uncertainties do too: for example, in Fig. \ref{fig:oed-introduction}(b) note that compared to the blue data and parameter estimates, the more accurate green data measurement produces slightly larger parameter uncertainties due to the lower gradient of \F encountered during back-projection. In this case, the range of parameter values that we expect might occur (our prior information about the parameters) thus influences which design is optimal. An average slope over the range of possible model parameter values can thus be used as a design quality metric for slightly non-linear forward functions. Since the range of parameter values that might be encountered is described by the Bayesian prior probability distribution, optimising this quality metric is typically referred to as \textit{Bayesian optimal experimental design} in statistical literature. However, this terminology is misleading, because this approach only optimises an average of a design quality measure that only accounts for physics that is linearised around each parameter value. It would therefore be more appropriate to refer to these as pseudo-Bayesian or linearised design methods \citep{Pronzato1985-fs,Chaloner1995-bs,Fedorov1997-za, Winterfors2012-uc, Ryan2016-yo}.

The simple averaging method described above breaks down for generally non-linear functions, especially for those that are multi-modal (have multiple distinct peaks and troughs). An example of such a function and how it affects the mapping from data to model parameter space is given in Fig. \ref{fig:oed-introduction}(c). This shows that a single datum may be consistent with different distinct regions of model parameter space far removed from each other (green). Even a datum whose mean measurement value is consistent with only a single model can also map to a range of distinctly different model parameter values due to its measurement uncertainty around the mean (blue). If we are to define a quality measure that describes the aspects of any experimental design for a fully non-linear forward function, clearly, it must depend on all possible models that explain the data. Since the set of parameter values consistent with the data is then non-unique, disjoint, and may have a varying probability of being true given the data, we describe the set of values by the Bayesian posterior pdf. The most commonly used design quality metric in substantially non-linear problems is the \textit{expected information gain} (\EIG) \citep{Lindley1956-lu}, which will be introduced in section \ref{sec:EIG_est}.

This work aims to introduce a set of Bayesian optimal experimental design methods that are novel to Geophysics, and which calculate the \EIG in the context of geophysical applications with generally linear or nonlinear forward functions. We illustrate their relative merits and computational costs in the context of three representative Geophysical experiments: (i) locating seismic sources based on P and S wave arrival times, (ii) assessing the contrast in seismic velocity across a stratal interface given measurements of the amplitudes of waves reflected from that interface, and (iii) designing a survey to best constrain CO2 saturation in a subsurface storage scenario. While these examples concern seismic waves, they are representative of design problems for the location of other source types \citep{Kim2014-ma, Lugrin2014-ii} and for reflections of other wave types \citep{Tarantola1984-as, Hunziker2016-pp}, since elastic waves exhibit intermediate physical complexity between acoustic and electromagnetic waves. We also demonstrate that optimal designs may change substantially depending on which question about the subsurface we wish to answer \citep{Arnold2018-wx}. These results and illustrations show that optimal design methods might usefully be adopted more widely in Geophysics as they are in other scientifically advanced disciplines, and they allow practitioners to make more informed choices between the various methods for their particular applications.

%% file: BED.tex
The Bayesian posterior distribution is denoted $p(\mitbf{m} \, | \, \mitbf{d})$ and describes the state of knowledge post-experiment. It can be expressed using Bayes rule as
\begin{align}
  p(\mitbf{m} \, | \, \mitbf{d}) = \frac{p(\mitbf{m}) p(\mitbf{d} \, | \, \mitbf{m})}{\int_{\mathbb{M}} p\left(\mitbf{m}^{\prime}\right) p\left(\mitbf{d} \, | \, \mitbf{m}^{\prime}\right) d \mitbf{m}^{\prime}} = \frac{p(\mitbf{m}) p(\mitbf{d} \, | \, \mitbf{m})}{p(\mitbf{d})} \label{eqn:bayes_general}
\end{align}
where \prior is the prior pdf, $p(\mitbf{d} \, | \, \mitbf{m})$ is called the likelihood function and $p(\mitbf{d})$ is called the evidence. The prior and posterior pdf's are defined over the model parameter space $\mathbb{M}$, while the likelihood and evidence are defined over data space $\mathbb{D}$. The dimensionality of these spaces is a primary factor in the computational complexity of both inference and design problems.

An intuitive example of the model parameter space is the three-dimensional space for seismic source (earthquake) location, with an optional fourth dimension for the origin time. A corresponding data space might be the $n$-dimensional space of arrival times of first-arriving waves detected at each of $n$ receivers. Alternatively, we might decide to pick both P and S wave arrival times, in which case the data space might consist of these data at $n/2$ receivers. This example thus illustrates that the choice of data processing algorithms may change the data space substantially, and so may be a primary element of any experimental design.

Equation \eqref{eqn:bayes_general} can be used to characterise the solution to many geophysical parameter estimation problems. Since the interest of this paper lies in experimental design, it is necessary to include a variable describing the design $\xi \in \Xi$, where $\Xi$ denotes the space of all potential experimental setups. Provided that the experimental design does not impose a change in the parametrisation of the model, the prior distribution \prior is usually not affected by a change in design. The design influences the solution to any inference problem through the likelihood \likelihood, which in turn affects the evidence. The posterior of the model parameters given an observation and experimental design is therefore 
\begin{align}
p(\mitbf{m} \, | \, \mitbf{d}, \xi)=\frac{p(\mitbf{m}) p(\mitbf{d} \, | \, \mitbf{m}, \xi)}{\int_{\mathbb{M}} p\left(\mitbf{m}^{\prime}\right) p\left(\mitbf{d} \, | \, \mitbf{m}^{\prime}, \xi \right) d \mitbf{m}^{\prime}} =\frac{p(\mitbf{m}) p(\mitbf{d} \, | \, \mitbf{m}, \xi)}{p(\mitbf{d} \, | \, \xi)} \label{eqn:bayes_design_specific}
\end{align}
The evidence \evidence acts as a normalising factor for the resulting posterior pdf and is sometimes not calculated explicitly when solving inference problems - for example, commonly used Markov chain Monte Carlo Methods are designed to avoid its calculation \citep{Mosegaard1995-pi}. However, many non-linear OED algorithms depend on \evidence. For nonlinear problems, calculating either \posterior or \evidence generally requires many evaluations of \likelihood to estimate the integral expression in Equation \eqref{eqn:bayes_design_specific}, and each evaluation of \likelihood requires a computation of the forward function. Therefore, the tractability of design problems depends on the complexity of \F and the number of evaluations required to estimate this integral, unless a way to avoid its evaluation or a sufficiently accurate approximation is found. For a general, yet also detailed mathematical treatment of Bayesian inference and for additional examples, we refer readers to \citet{Tarantola2005-gj}.

Generally, we aim to optimise experiments such that they maximise information in the posterior distribution, within bounds imposed by practical constraints on the cost of performing the experiment. We therefore need a metric that quantifies the information embodied within any probability distribution. Shannon information \citep{Shannon1948-of} is an intuitive measure of information with several beneficial properties (\eg linear additivity of information from independent sources). The Shannon information $I[\cdot]$ described by an arbitrary continuous probability density function $p(x)$ is defined as
\begin{align}
    \mo{I} \left[p(x) \right] = \mathbb{E}_{p(x)} \left[ \log _{b}\left(p(x) \right) \right] = \int_{\mathcal{X}} p(x) \log _{b}(p(x)) d x \label{eqn:information}
\end{align}
where $x\in \mathcal{X}$ is a random variable distributed according to $p(x)$ and $\mathbb{E}_{p(x)}$ is the expectation with respect to $p(x)$, which is defined by the right-most expression. Depending on the context, information is also often expressed as the negative of the entropy $\mo{H}$, $\mo{I} \left[p(x) \right] = - \mo{H}[p(x)]$, where entropy $\mo{H}$ is defined to be the negative of either expression on the right of Equation \eqref{eqn:information}. This absolute information measure can be extended to the relative information content of one pdf relative to another, also called the Kullback-Leibler (KL) divergence \citep{Kullback1951-pq}
\begin{align}
    \text{KL}(P||Q) = \int_\mathcal{X} p(x) \log_b\left( \frac{p(x)}{q(x)} \right) d x
\end{align}
For further information on the properties of information, the reader is referred to \citet{Cover2006-bi}.

Following \citet{Lindley1956-lu}, we now define the information gain (\IG) about \m obtained by recording data \d using experimental design $\xi$, to be the difference between the posterior state of information and the information about \m in the prior pdf:
\begin{align}
    \mo{IG}(\xi, \mitbf{d}) & = \mo{I} \left[p(\mitbf{m} \, | \, \mitbf{d}, \xi)] - \mo{I}[p({\mitbf{m}}) \right] \label{eqn:IG_short}\\
                         & =          \mathbb{E}_{p(\mitbf{m} \, | \, \mitbf{d}, \xi)}[\log(p(\mitbf{m} \, | \, \mitbf{d}, \xi))]-\mathbb{E}_{p(\mitbf{m})}[\log(p(\mitbf{m}))]\label{eqn:IG_long}
\end{align}
where Equation \ref{eqn:IG_long} is obtained by substituting Equation \ref{eqn:information} into \ref{eqn:IG_short}. The \IG depends on \d, which is not available during the design of an experiment. However, the evidence \evidence provides the probability of observing any particular value for \d, given our prior state of information about the parameters, and the forward function and measurement uncertainties which are both included within the likelihood (Equation \eqref{eqn:bayes_design_specific}). The so-called \textit{expected information gain} (\EIG) is therefore defined as the expectation of \IG over \evidence, giving
\begin{align}
    \mo{EIG}(\xi) = \mathbb{E}_{p(\mitbf{d} \, | \, \xi)} \left[\mo{I}[p(\mitbf{m} \, | \, \mitbf{d}, \xi)] - \mo{I}[p({\mitbf{m}})] \right] \label{eqn:EIG_def_model}
\end{align}
This criterion depends only on \design and is used in this work to assess the quality of any experimental design \design prior to the collection of data.

While Equation \eqref{eqn:EIG_def_model} is perhaps the most intuitive way to express the EIG, its value is often calculated using other expressions. Rewriting Equation \eqref{eqn:EIG_def_model} using Equation \ref{eqn:bayes_general} we obtain:
\begin{align} 
    \operatorname{EIG}(\xi) & = \mathbb{E}_{p(\mitbf{d}, \mitbf{m} \, | \, \xi)}\left[\log \frac{p(\mitbf{m} \, | \, \mitbf{d}, \xi)}{p(\mitbf{m})}\right] \label{eqn:EIG_def_detail_model}\\
                            & = \mathbb{E}_{p(\mitbf{d}, \mitbf{m} \, | \, \xi)}\left[\log \frac{p(\mitbf{d}, \mitbf{m} \, | \, \xi)}{p(\mitbf{m}) p(\mitbf{d} \, | \, \xi)}\right]\\
                            & = \mathbb{E}_{p(\mitbf{d}, \mitbf{m} \, | \, \xi)}\left[\log \frac{p(\mitbf{d} \, | \, \mitbf{m}, \xi)}{p(\mitbf{d} \, | \, \xi)}\right] \label{eqn:EIG_def_detail_data}
\end{align}
In these expressions, the information is written out explicitly, and all three lines are equivalent and can be derived by repeated use of Bayes theorem. Both Equations \eqref{eqn:EIG_def_model} and \eqref{eqn:EIG_def_detail_model} are written only in terms of pdfs in the model parameter space, meaning they assign probabilities to each vector of model parameter values (\eg earthquake locations in the example introduced above). This means that evaluating the \EIG in this form requires that we solve an inverse problem to obtain the posterior distribution \posterior for each data vector \d (or a representative subset of data samples) that is likely to be observed according to \evidence. For many problems, solving just one inverse problem is a considerable computational effort, which often makes this approach impractical. Using Equation \eqref{eqn:EIG_def_detail_data}, on the other hand, the \EIG is expressed only in terms of distributions in data space, \likelihood and \evidence, as
\begin{align}
    \mo{EIG}(\xi) \triangleq \mathbb{E}_{p(\mitbf{m})}\left[ \operatorname{I}[p(\mitbf{d} \, | \, \mitbf{m}, \xi)] - \mo{I}[p(\mitbf{d} \, | \, \xi)] \right] \label{eqn:EIG_def_data}
\end{align}
making it possible to design experiments without explicitly solving inverse problems. In this formulation, the main difficulty lies in estimating $\mo{I}\left[p(\mitbf{d} \, | \, \xi)\right]$. Especially for high-dimensional data spaces (\eg arrival times recorded by many receivers), this is challenging so for problems with low model parameter space dimensions and high data space dimensions, it may still be beneficial to calculate the \EIG in the model parameter space (Equation \eqref{eqn:EIG_def_model}). Therefore, both model parameter and data space approaches are analysed in more detail herein.

Using either expression for the \EIG as our design metric, the best design is expressed mathematically as
\begin{align} 
    \xi^{*} &= \underset{\xi \in \Xi}{\arg \max } \; \mo{EIG} (\xi) \label{eqn:eig_optimization}
\end{align}
where $\Xi$ is the set of all possible experimental designs. Ideally, this optimisation is global, but in many cases a greedy (local) optimisation algorithm must be used to make the optimisation computationally tractable. Such algorithms nevertheless usually provide a significant improvement over non-optimised experiments. Further details on the optimisation process are given in section \ref{sec:EIG_opt}.

\subsection{Comments on Linear Design Theory}

  This paper focuses on algorithms applicable to the complex forward functions that occur in nature, so methods designed specifically for linear or linearised models will not be covered in detail. However, because linear(-ised) forward functions are still widely used in Geophysics we provide a brief description of the most popular methods, and an explanation of why linear design methods are unlikely to work well in most geophysical problems. For a more detailed treatment of linear experimental design measures, the reader is referred to \citet{Atkinson1992-sw} for a general overview, and to \citet{Curtis1999-hf} for an overview of most linear design measures used in geophysics.

  By definition, in linear models the model-data relation can be written
  \begin{equation}
      \mitbf{d} = \mitbf{A} \mitbf{m} + \mitbf{\varepsilon} \label{eqn:linear_relation}
  \end{equation}
  where $\mitbf{A}$ describes the linear relations between \d and \m and is often referred to as the design matrix, any constant offset has been absorbed into \d, and $\mitbf{\varepsilon}$ is a vector of random data noise. To find the inverse relation in the case of an overdetermined problem, Equation \eqref{eqn:linear_relation} is typically premultiplied by $\mitbf{A}^T$ to obtain a square matrix which can then be inverted to estimate a model $\mitbf{m}_0$ that gives the best fit in the least squares sense
  \begin{align}
      \mitbf{m}_\text{inv} = \left( \mitbf{A}^T\mitbf{A} \right)^{-1} \mitbf{A}^T (\mitbf{d}_\text{obs} - \mitbf{\varepsilon})
  \end{align}
  This inversion is highly sensitive to the eigenvalue spectrum ($N$ eigenvalues $\{\lambda_i; i=1,...,N\}$ in order of decreasing magnitude) of matrix $\mitbf{L} = \mitbf{A}^T \mitbf{A}$, because low eigenvalues amplify measurement errors by a factor $1/\lambda_i$. For this reason, a generalised (regularised) inverse $\mitbf{L}^{\dagger}$ is used in practice to avoid the instability of $\mitbf{L}^{-1}$, and this is obligatory in underdetermined problems \citep{Menke2018-ji}. Linear experimental design methods maximise some measure of the size of the eigenvalues of $\mitbf{L}$ to allow the best possible approximation of $\mitbf{L}^{-1}$ to be used, and thus to minimise the amplification of errors \mitbf{\varepsilon} in the solution. This leads to the so-called alphabetic design criteria \citep{Box1959-ks, Atkinson1992-sw}, which have well-studied properties \citep{Kiefer1959-rs}. Descriptions of the most commonly used alphabetic criteria follow.

  \paragraph*{D-optimality}

    The most popular linear design criteria is the determinant- or D-criterion
    
    \begin{align}
      \det{\mitbf{L}} = \prod_{i=1}^N \lambda_i \label{eqn:d-optimality}
    \end{align}
    which is inversely proportional to the determinant of model posterior covariance matrix given Gaussian data noise. The product in equation \eqref{eqn:d-optimality} makes the D criterion very sensitive to small eigenvalues and inapplicable for underdetermined problems. Adapting the measure to include a lower eigenvalue threshold and a penalty term enables the D-criterion to be used in such situations (for details see appendix of \citet{Curtis1999-hf}, first introduced to geophysics by \citet{Maurer1998-vd}). Calculating the determinant of a matrix makes this criterion expensive to evaluate.

  \paragraph*{A-optimality}

    Another popular choice in linear experimental design is the A-criterion

    \begin{align}
      \mo{Tr}{\mitbf{L}} = \sum_{i=1}^N \lambda_i
    \end{align}
    which represents the area under the eigenvalue spectrum. Maximising this measure typically results in designs that maximise a subset of large eigenvalues at the expense of small ones; this in turn focuses information on linear combinations of parameter values defined by the eigenvectors corresponding to those large eigenvalues \citep{Curtis1999-hf}. In other words, in linear problems, an A-optimal model would achieve highly reliable information about a limited number of linear combinations of model parameters. An alternative formulation that avoids this limitation is normalised A-optimality, in which the trace of $\mitbf{L}^{-1}$ is divided by the largest eigenvalue $\lambda_1$. The optimality criterion then becomes the area under the normalised eigenvalue curve, which assigns more equal importance across the range of eigenvalues.
    
    Both A-optimality measures quantify properties of matrix $\mitbf{L}^{-1}$ as a whole. If only a subset of the model parameters (or a subset of linear parameter combinations) is of interest, focused A-optimality can be used by weighting the squared sum of eigenvalues by the squared dot product of the corresponding eigenvector and the basis vectors of the model region of interest \citep{Curtis1999-mg}.
    The main advantage of A-optimality over D-optimality is that the trace of $\mitbf{L}$ can be estimated cheaply as it does not require the full eigenvalue spectra to be calculated, and the maximum eigenvalue used as a normalising factor in normalised A-optimality can be estimated inexpensively using the power method \citep{Curtis1997-iq}.

  \paragraph*{E-optimality}
    If the eigenvalues $\lambda_i$ of $\mitbf{L}$ are ordered by decreasing magnitude ($\lambda_1$ largest, $\lambda_N$ smallest), the E-optimality condition maximises the smallest eigenvalue $\lambda_N$. This can also be generalised to maximising a specific eigenvalue $\lambda_k$, so $\lambda_k$ is the design criterion. In some practical applications (\eg \cite{Curtis1999-mg}), the logarithm of the criteria is taken for computational stability which does not influence the resulting optimal design since the logarithm is a monotonic function.

  \paragraph*{Relationship between alphabetic criteria and EIG}

    The alphabetic criteria can be related to the \EIG for Bayesian linear models. If the prior pdf is Gaussian $\mitbf{m} = \mathcal{N}\left(m_0, \mitbf{\Sigma}_0\right)$, then for linear models the posterior pdf is proportional to $\left( \mitbf{L} + \mitbf{\Sigma}_0^{-1}\right)^{-1}$ and its information content is independent of the observed data. Making use of the analytical form of the entropy of a Gaussian, the \EIG can be expressed as
    \begin{align}
      \mo{EIG}(\mitbf{A}) = \frac{1}{2} \log{|\mitbf{\Sigma}_0|} - \frac{1}{2} \log{|\mitbf{L} + \mitbf{\Sigma}_0^{-1}|} + C = \log{|\mitbf{L}|} + C^{'} \label{eqn:EIG_linear_gaussian}
    \end{align} 
    where C and C' are constants. Therefore maximisation of $|\mitbf{L}|$ in the D-criterion results in the same design as would be obtained by maximising the \EIG.

  \paragraph*{Linearised Models}

    Extensive work has been made to extend the alphabetic-criteria developed for linear problems to non-linear problems \citep{Tsutakawa1972-kv, Chaloner1995-bs}. For this, a linearised version of Equation \eqref{eqn:general_relation} is used
    \begin{align}
      \mitbf{d} = \mitbf{A}_{m_0} \mitbf{m} + \mitbf{\varepsilon} \label{eqn:linearised_relation}
    \end{align}
    where $\mitbf{A}_{\mitbf{m}_0}$ is the Jacobian matrix of partial derivatives of $F(\mitbf{m})$ with respect to $\mitbf{m}$, evaluated at (linearised around) reference model $\mitbf{m}_0$. The physical relationships between \d and \m are thus only approximate, and are only accurate in some locality of $\mitbf{m}_0$. Consequently, the design that results from optimising this relationship would no longer be independent of the reference model - which in turn should be free to vary according to the prior pdf. By taking the expectation over reference models, a Bayesian, linearised version, of the D-optimality criterion is given by \citet{Chaloner1995-bs}:
    \begin{align}
      \mathbb{E}_{p(\mitbf{m})}\left[ \log \det{\mitbf{\tilde{L}_{m}}} \right] \label{eqn:linearised_D_criterion}
    \end{align}
    where $\mitbf{\tilde{L}_{m}} = \mitbf{A}_{m}^T \mitbf{A}_{m}$. The other alphabetic criteria can be extended to linearised problems using expectations in the same way.
    
    The problems with this approach are apparent in Fig. \ref{fig:oed-introduction}: an averaging approach may work intuitively for non-linear but approximately monotonic functions Fig. \ref{fig:oed-introduction}(b)  because the resulting estimate of design quality is inversely related to the expected uncertainty of the solution approximated locally by linearised physics. However, as shown in Fig. \ref{fig:oed-introduction}(c), in the presence of multimodality an approach based on derivatives breaks down, because the solution uncertainty approximated by the gradient of \F only approximates one of the modes of the posterior pdf over parameter values that are consistent with the data \citep{Winterfors2012-uc}. This is clear from Equation \ref{eqn:linearised_D_criterion}, which shows that around each prior value of the parameter space at which $\mitbf{{L}_{m}}$ is evaluated, the posterior solution accounted for in Equation \ref{eqn:EIG_linear_gaussian} consists of a single Gaussian. Therefore, even the linearised uncertainty estimate represented by the term in brackets in Equation \ref{eqn:linearised_D_criterion} is substantially incorrect.

%% file: EIG_estimation.tex
The mathematical formulation of the \EIG is straightforward. However, its evaluation is not, as neither \evidence nor \posterior are typically known in closed form, and both \d and \m over which each is defined can have many dimensions. The following sections present approaches to computing (an approximation of) the \EIG corresponding to any given design.

\begin{figure}
  \centering
  \includegraphics[width=0.5 \textwidth]{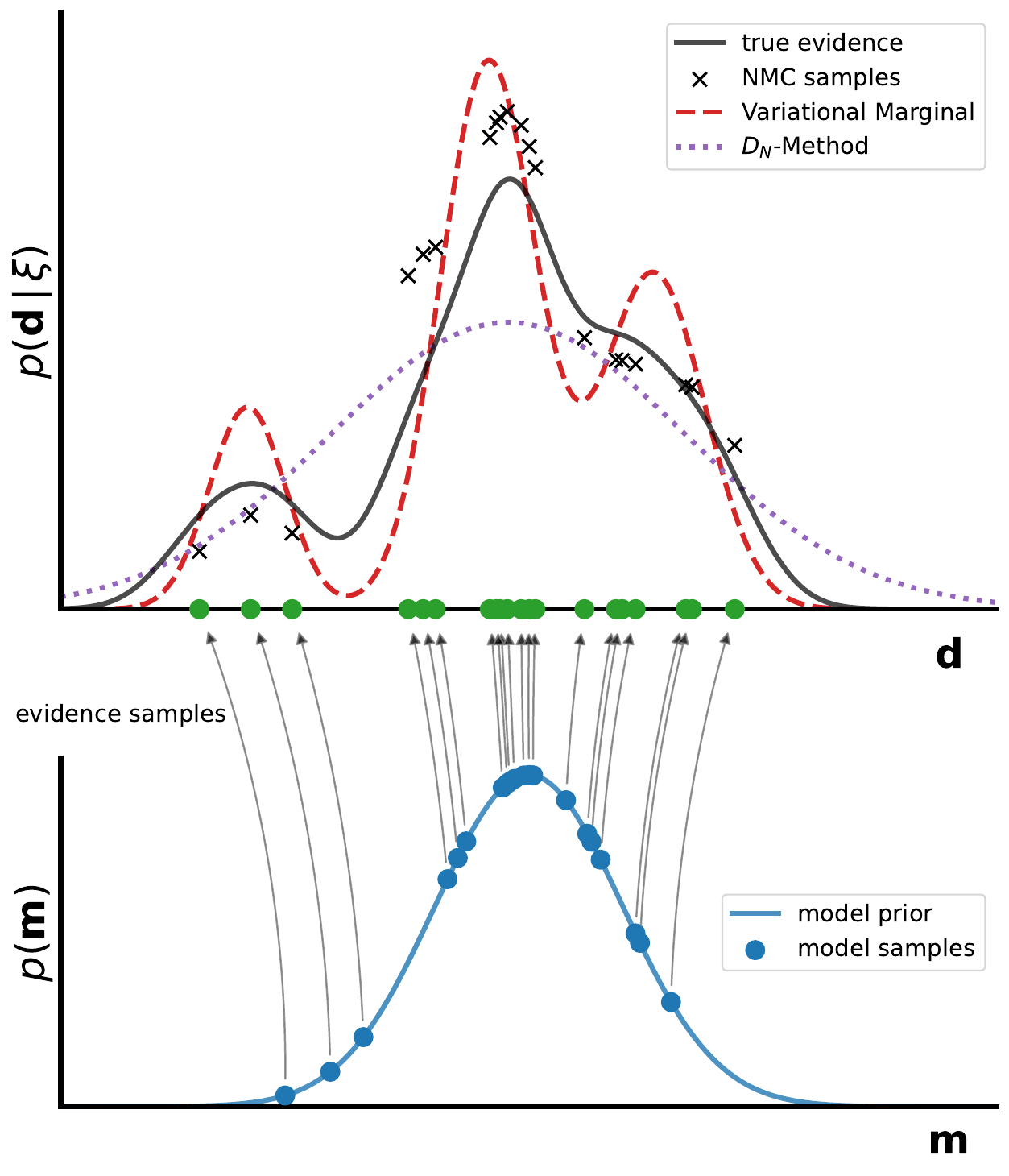}
  \caption{
    Illustration of expected information gain estimation methods in data space. Samples in model parameter space (blue dots) are generated from the prior pdf (blue curve). The likelihood (and the underlying forward model evaluation) can be used to generate samples from the evidence in data space (green dots) using the model space samples. Details of how \evidence is estimated using various methods (crosses, and blue and red curves in top graph) described in Section \ref{sec:EIG_est}. NMC denotes Nested Monte Carlo.
  }\label{fig:dataspace_methods}
\end{figure}

The basis for the following methods is to take samples $\mitbf{m}_i$ of the prior pdf (bottom of Fig. \ref{fig:dataspace_methods}) and to project them into the data space by evaluating the forward function and taking samples $\mitbf{d}_i$ of the likelihoods $p(\mitbf{d}_i \, | \, \mitbf{m}_i, \xi)$ (forward model evaluation of prior samples is used to construct likelihoods). This procedure provides samples of the evidence (green dots in Fig. \ref{fig:dataspace_methods}) but no probability values for those points in data space because the explicit evaluation of \evidence requires to solve the integral in Equation \ref{eqn:bayes_design_specific} which is over the whole model space. The nested Monte-Carlo, variational marginal, and $D_N$ methods use different approaches to estimate $p(\mitbf{d}_i \, | \, \xi)$ needed to estimate the information described by the evidence and are therefore referred to as \textit{data space methods}.

\subsection{Monte Carlo Methods}

  The most straightforward and robust, but computationally expensive, way of evaluating Equation \eqref{eqn:EIG_def_data} is to deploy a naive nested Monte-Carlo (MC) approach as introduced by \citet{Ryan2003-qp} and \citet{Myung2013-vp} (further analysed in detail by \citet{Vincent2017-fs} and \citet{Rainforth2017-rp}).

  The so-called nested Monte-Carlo (NMC) \EIG estimator is defined as
  \begin{align}
      \mo{EIG}_{\mathrm{NMC}}(\xi) = \frac{1}{N} \sum_{i=1}^{N} \log \frac{p\left(\mitbf{d}_{i} \, | \, \mitbf{m}_{i, 0}, \xi\right)}{\frac{1}{M} \sum_{j=1}^{M} p\left(\mitbf{d}_{i} \, | \, \mitbf{m}_{i, j}, \xi\right)} \label{eqn:NMC}
  \end{align} 
  where $\mitbf{m}_{i, j} \sim p(\mitbf{m})$ and $\mitbf{d}_{i} \sim p \left( \mitbf{d} | \left[\mitbf{m}=\mitbf{m}_{i, 0} \right], \xi\right)$. The nested loop results in a slightly unusual sampling notation. A $N \times (M+1)$ array of prior pdf samples $\mitbf{m}_{i, j}$ is generated from \prior. The first row in this square is then referred to by $\mitbf{m}_{i, 0}$, which means the outer loop only uses samples of this first row while the inner loop uses a different column $j$ (excluding the first element) of samples for each step of the outer loop. 
  
  Pointwise estimates of the evidence $p\left(\mitbf{d}_{i} \, | \,\xi \right)$ at $N$ points $\mitbf{d}_{i}$ in data space are calculated using the average of the $M$ likelihood functions $\frac{1}{M} \sum_{j=1}^{M} p\left(\mitbf{d}_{i} \, | \, \mitbf{m}_{i, j}, \xi \right)$  (inner loop of Equation \eqref{eqn:NMC}), which estimate how likely it is that one would observe the datum $\mitbf{d}_{i}$. This results in a set of $N$ points with an assigned probability in data space (grey crosses in Fig. \ref{fig:dataspace_methods}). As these points are sampled from the prior, they can then be used to calculate an MC estimate of $\mo{I}\left[ p(\mitbf{d \, | \, \xi})\right]$ by $\frac{1}{N} \sum_{j=1}^{N} p\left(\mitbf{d}_{i} \, | \,\xi \right)$. The likelihood term $p\left(\mitbf{d}_{i} \, | \, \mitbf{m}_{i, 0}, \xi\right)$ of Equation \eqref{eqn:NMC} takes the influence of the data noise into account which might change for different designs \design (\eg receiver positions close to noise sources may produce larger measurement uncertainties than those located in quiet areas).
  
  The estimator $\mo{EIG}_{\mathrm{NMC}}$ has a computational cost of $T=\mathcal{O}(N M)$ with an RMSE (root-mean-square error) convergence rate of $\mathcal{O}(N^{-1/2} M^{-1})$ which means it is asymptotically optimal to set $M \propto \sqrt{N}$. The number of inner samples $M$ controls the bias, while the number of outer samples $N$ controls the variance of this quality estimate \citep{Huan2013-nf}. The computational cost of the NMC approach can be reduced considerably if the same samples of the inner loop ($M$) are reused for each iteration of the outer loop as demonstrated by \citet{Huan2013-nf}, \citet{Qiang2022-ow} and \cite{Zhang2015-sp}, resulting in a computational cost of $T=\mathcal{O}(N+M)$ total samples. This reuse of samples increases the bias in the \EIG estimate, but if the bias is stationary this would not affect the subsequent design optimisation. To our knowledge no useful bounds on the size of the bias, nor practically implementable conditions that guarantee stationarity for particular problems are available. The $\mo{EIG}_{\mathrm{NMC}}$ estimate is an upper bound and will therefore always be larger than the true \EIG \citep{Foster2019-rx}.

\subsection{Maximum Entropy Method}

  If the likelihood is independent of the design, $p(\mitbf{d} \, | \, \mitbf{m}, \xi) = p(\mitbf{d} \, | \, \mitbf{m})$, meaning that the measurement uncertainty on each datum does not vary with the design (for example, with receiver location). Then Equation \eqref{eqn:EIG_def_data} is equivalent to
  \begin{align}
      \mo{EIG}_\mathrm{ME}(\xi) = -\mo{I}\left[p(\mitbf{d} \, | \, \xi)\right] + C
  \end{align}
  Maximising the EIG is then equivalent to maximising the entropy (negative information) of the evidence, resulting in so-called maximum entropy design \citep{Shewry1987-wu}. This only slightly lowers the cost compared to the NMC method, since the main cost in evaluating Equation \eqref{eqn:EIG_def_data} is the estimation of $\mo{I}\left[p(\mitbf{d} \, | \, \xi)\right]$ which can be taken out of the expectation due to its independence of \m. This cost is the same in both the NMC and the maximum entropy method. $\mo{EIG}_\mathrm{ME}$ can therefore be estimated either by using a similar nested Monte Carlo loop as for NMC or using specific methods to calculate the entropy of a set of samples, such as k-d-partitioning \citep{Stowell2009-iu, Bloem2020-gp}.

  Especially in cases with a large number of data dimensions (\eg observations made using many receivers), both the NMC and the maximum entropy estimator, while consistent, will converge prohibitively slowly to be practical for most geophysical applications. Perhaps luckily then, experimental design is often particularly effective for experiments with restricted resources which may have fewer observations and hence data space dimensions, making MC methods useful in many cases. MC methods are also necessary for benchmarking algorithms that use approximations introduced in the following sections.

\subsection{Variational Methods}

  The bottleneck in evaluating Equations \eqref{eqn:EIG_def_model} and \eqref{eqn:EIG_def_data} occurs in the estimation of \posterior and \evidence, respectively. The main inefficiency in the NMC method arises because the integrand in equation \eqref{eqn:NMC} is estimated separately for each \d. Instead of evaluating this integrand directly, \citet{Foster2019-rx} proposed to learn a \textit{variational} (closed form, or otherwise analytic) approximation to either \posterior or \evidence. 
  
  Suppose the construction of this functional approximation requires $M$ samples. In that case, the total computational cost is on the order of $\mathcal{O}(N + M)$, which may be a substantial reduction compared to the NMC method, so we now introduce variational approaches to \EIG estimation.
      
  \subsubsection{Variational Marginal Method}

    The \textit{variational marginal method} operates in data space by finding a variational estimator $q_{m}\left(\mitbf{d} \, | \, \xi \right)$ that approximates the evidence \evidence. The evidence is in fact the data space marginal posterior pdf, hence the name of this design method. Instead of evaluating the marginal density \evidence for each of the $N$ data samples $d_n$, the idea is to train a variational functional emulator $q_m$ using $M$ samples. Once $q_m$ is available, the \EIG can be approximated as
    \begin{align}
      \operatorname{EIG}_\text{marg}(\xi) = \frac{1}{N} \sum_{n=1}^{N} \log \frac{p\left(\mitbf{d}_{n} \, | \, \mitbf{m}_{n}, \xi\right)}{q_{m}\left(\mitbf{d}_{n} \, | \, \xi\right)} \label{eqn:var_mar}
    \end{align}
    where $\mitbf{m}_{i} \sim p(\mitbf{m})$ and $\mitbf{d}_{i} \sim p \left(\mitbf{d} | \left[\mitbf{m}=\mitbf{m}_{i} \right], \xi\right)$. 
    
    The variational approximator $q_m(\mitbf{d} \, | \, \xi)$ is found by first introducing a variational family $q_m(\mitbf{d} \, | \, \xi, \boldsymbol{\phi})$ (\eg the family of multivariate Gaussians) with parameters $\boldsymbol{\phi}$ (\eg describing mean and covariance) that parametrise possible forms of $q_m$. The key is to find parameters $\boldsymbol{\phi}$  that provide the best possible approximation $q_m$ to \evidence on average for any \d. It can be shown (see the appendix of \citet{Foster2019-rx}) that this formulation yields an upper bound to the true \EIG, which is tight strictly if and only if $q_m(\mitbf{d} \, | \, \xi) = p(\mitbf{d} \, | \, \xi)$. The optimal parameters $\boldsymbol{\phi}^*$ can therefore be found by using stochastic gradient descent (SGD) \citep{Robbins1951-tw} to solve the following optimisation problem: 
    \begin{align}
      \boldsymbol{\phi}^{*} 
      &=\underset{\boldsymbol{\phi}}{\arg \max } \left\{ \mathbb{E}_{p(\mitbf{d}, \mitbf{m} \, | \, \xi)}\left[ \log q_{m}(\mitbf{d} \, | \, \xi, \boldsymbol{\phi})\right] - \underbrace{ \mathrm{I} \left[p(\mitbf{d} \, | \, \mitbf{m}, \xi)\right]}_\text{constant} \right\} \label{eqn:var_mar_estimation}
    \end{align}
    which means that we want to find the variational family for which the evidence samples have the highest expected probability of being observed. 
    
    The maximisation of the excepted likelihood of $q_m$ will, in turn, minimise Equation \eqref{eqn:var_mar} and cause $\operatorname{EIG}_\text{marg}(\xi)$ to converge towards the true $\operatorname{EIG}(\xi)$. The tightness of the bound is determined by how well  $q_m(\mitbf{d} \, | \, \xi, \boldsymbol{\phi}^{*})$ represents the true evidence.
    
    The variational family for the variational marginal method used in this study is a Gaussian mixture model - a sum of Gaussians \citep{Bishop2006-yk}. The red curve in Fig. \ref{fig:dataspace_methods} illustrates an example variational marginal pdf. The density of the evidence samples (green) is used to learn a function, here a sum of Gaussians, which can be evaluated straightforwardly to approximate the probability of observing points in data space. These probabilities are then used to calculate the \EIG. We have also successfully applied normalising flows \citep{Tabak2013-ib, Dinh2014-xp, Rezende2015-ql, Zhao2020-nf, Durkan2019-dl} to construct the above function, but they are omitted in this paper for brevity.

  \subsubsection{Variational Posterior Method}\label{sec:var_post}

    Training an approximator to \evidence might be disadvantageous if the number of data dimensions is high and the number of model dimensions is low (\eg when designing a seismic source location experiment with many receivers recording travel time measurements). Equation \eqref{eqn:EIG_def_model} allows us instead to calculate the \EIG in the model domain by finding an approximator to \posterior. This is achieved by finding a variational function $q_p(\mitbf{m} \, | \, \mitbf{d}, \xi)$ to represent the posterior pdf given an observed datum \d measured under design \design. The \EIG can then be approximated by the \textit{variational posterior method} as
    \begin{align}
      \operatorname{EIG}_\text{post}(\xi) = \frac{1}{N} \sum_{n=1}^{N} \log \frac{q_{p}\left(\mitbf{m}_{n} \, | \, \mitbf{d}_{n}, \xi \right)}{p\left(\mitbf{m}_{n}\right)} \label{eqn:var_post}
    \end{align}
    where $\mitbf{m}_{i} \sim p(\mitbf{m})$ and $\mitbf{d}_{i} \sim p \left(\mitbf{d} | \mitbf{m}=\mitbf{m}_{i}, \xi\right)$. 
    \begin{figure*}
      \centering
      \includegraphics[width=1.0\textwidth]{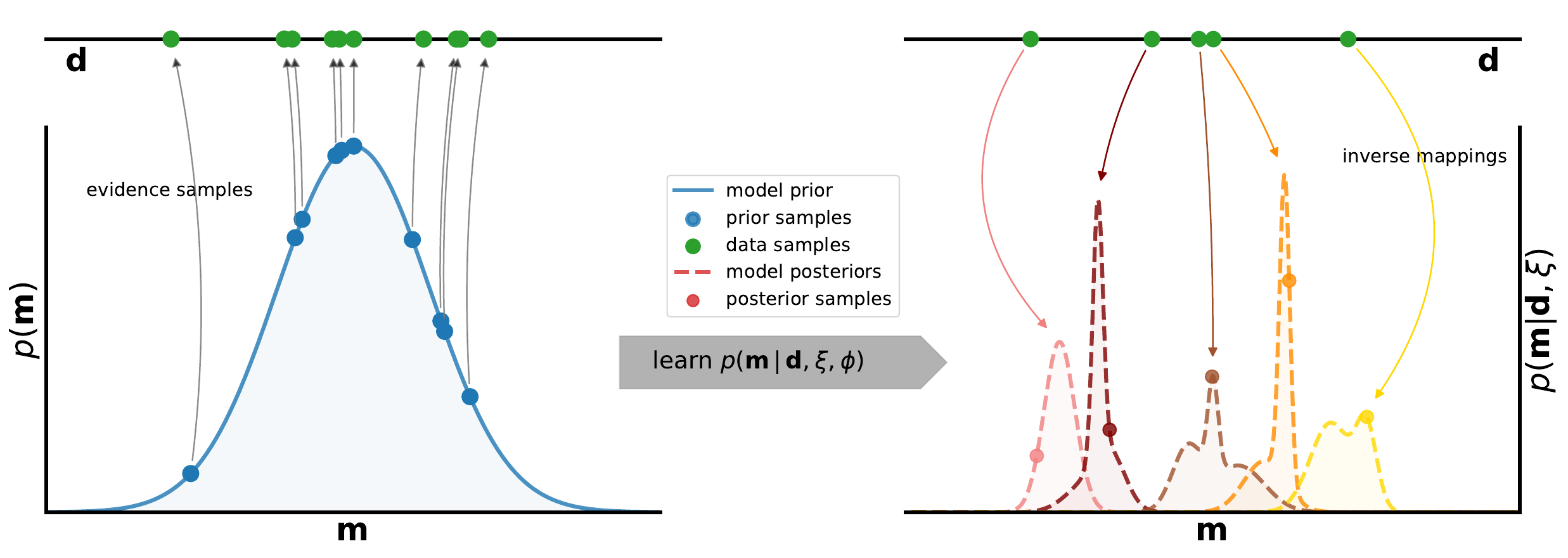}
      \caption{
        Illustrates the variational posterior method to estimate the expected information gain in model parameter space (left). The prior (blue curve) is used to generate samples in model parameter space (blue dots). The likelihood (and the underlying forward model evaluation) can be used to generate evidence samples in data space (green dots) using each model space sample. These pairs of model parameters and data samples can be used to learn a variational approximator $q_p(\mitbf{m} \, | \, \mitbf{d}, \xi, \boldsymbol{\phi})$ to \posterior. Thereafter, a different set of data samples is generated in the same way (right). The corresponding posterior pdf for every such sample can be estimated cheaply using mapping $q_p$.
      }\label{fig:modelspace_methods}
    \end{figure*}
    To evaluate this MC estimator, it is necessary to find the function $q_p(\mitbf{m} \, | \, \mitbf{d}, \xi)$. For this a family of variational distributions $q_p(\mitbf{m} \, | \, \mitbf{d}, \xi, \boldsymbol{\phi})$ parametrized by $\boldsymbol{\phi}$ is introduced. To learn $\boldsymbol{\phi}$ and therefore a function that is close to \posterior we can make use of the fact that $\operatorname{EIG}_\text{post}(\xi)$ is a lower bound  to the true \EIG, which is tight strictly if and only if $q_p(\mitbf{m} \, | \, \mitbf{d}, \xi, \boldsymbol{\phi}) = q(\mitbf{m} \, | \, \mitbf{d}, \xi)$. Maximising this lower bound to find the optimal choice $\boldsymbol{\phi}^{*}$ is equivalent to evaluating
    \begin{align}
      \boldsymbol{\phi}^{*}=\underset{\boldsymbol{\phi}}{\arg \max } \left\{ \mathbb{E}_{p(\mitbf{d}, \mitbf{m} \, | \, \xi)}\left[\log q_{p}(\mitbf{m} \, | \, \mitbf{d}, \xi, \boldsymbol{\phi})\right] - \underbrace{ \mathrm{I} \left[p(\mitbf{m})\right]}_\text{constant} \right\} \label{eqn:var_post_estimation}
    \end{align}
    which maximises the expected probability of observing the prior sample used to generate a data sample. The parameters $\boldsymbol{\phi}$ can be optimised using stochastic gradient descent. Then $q_p(\mitbf{m} \, | \, \mitbf{d}, \xi, \boldsymbol{\phi})$ results in a different pdf for each datum \d (orange pdfs on the right of Fig. \ref{fig:modelspace_methods}). The maximisation in Equation \eqref{eqn:var_post_estimation} is equivalent to minimising the KL divergence between $q_m(\mitbf{m} \, | \, \mitbf{d}, \xi, \boldsymbol{\phi})$ and \posterior. A considerable advantage of this approach is that it maximises the \EIG in the model parameter space, making it suited to design experiments with a high dimensional data space but a lower dimensional parameter space. 
    
    In this study, the variational family for the approximation of the variational posterior pdf is a mixture density network (MDN) \citep{Bishop1994-mf, Meier2009-qa}. We have also successfully applied conditional normalising flows \citep{Tabak2013-ib, Dinh2014-xp, Rezende2015-ql, Zhao2020-nf, Durkan2019-dl} but these are omitted in this paper for brevity.
    
    For both variational estimators $\operatorname{EIG}_\text{marg}$ and $\operatorname{EIG}_\text{post}$, the quality of the final result depends on how accurately $q_m$ or $q_p$ represents the true evidence or posterior. This depends on the flexibility (expressiveness) of the parametrisation of the two functional distributions, and on how well the gradient descent optimisation has converged.

\subsection{Covariance-based Methods}

  While variational methods can substantially reduce the computational cost of calculating the \EIG, their optimisation using stochastic gradient descent is still a significant computational expense. If the evidence can be approximated adequately by a multivariate Gaussian distribution $\mathcal{N}(\mu, \mitbf{\Sigma}_\text{evidence})$ with mean $\mu$ and covariance matrix $\mitbf{\Sigma}_\text{evidence}$, its information content can be calculated as a function of its covariance matrix by
  \begin{align}
    \mo{I} \left[ \mathcal{N}(\mu, \mitbf{\Sigma}_\text{evidence}) \right] = \frac{1}{k}(1+\ln(2\pi)) + \frac{1}{2} \ln \left( \left| \mitbf{\Sigma}_\text{evidence} \right| \right)  \label{eqn:information_multivariate_normal}
  \end{align}
  where $k$ is the dimensionality of the data space. Under the assumption of Gaussian data noise with covariance $\mitbf{\Sigma}_\text{data}$, the \EIG can then be expressed as 
  \begin{align}
    \operatorname{EIG}_{D_N} = \ln{\left| \mitbf{\Sigma}_\text{evidence} \right|} - \ln{\left| \mitbf{\Sigma}_\text{data} \right|} + C \label{eqn:DN}
  \end{align}
   the constant terms have been summarised in the constant $C$. This defines the so-called $D_N$ method, equivalent to the variational marginal method with the multivariate Gaussian variational family. It follows that it can be extended to non-Gaussian noise, taking only the numerator in equation \eqref{eqn:var_mar} and using it to replace $\ln{\left| (\mitbf{\Sigma}_\text{data}) \right|}$.

  This estimator was first introduced by \citet{Coles2011-dh} and applied by \citet{Rawlinson2012-cs, Coles2013-vw, Bloem2020-gp}. While Equation \eqref{eqn:DN} is only valid for a Gaussian distribution of the evidence, the $D_N$ criterion remains useful in many applications because maximising the covariance increases the spread in data space of data points corresponding to different models. Intuitively, the farther data from different models are spread apart, the easier it is to distinguish between models in the presence of data noise.

  Covariance-based measures can fail, most notably in the case of multimodality in the evidence, where the distance between modes does not necessarily influence the information but where a larger distance would lead to a higher covariance. Despite this limitation, the $D_N$ method appears to be essential in practice due to the efficiency of its evaluation and the small number of samples necessary to obtain a stable estimate of the \EIG. Figure (\ref{fig:dataspace_methods}) shows how the $D_N$ method compares to the NMC and the variational marginal method by showing the Gaussian (green) with the same mean and covariance as the evidence samples (green). It is not necessary to use this pdf to obtain sample probabilities as the information of a Gaussian is known in closed form (Equation \eqref{eqn:information_multivariate_normal}).

\subsection{Other Methods}

  This section has mainly focused on methods that have been used previously in geophysical applications, together with variational methods which have been developed in reasonable generality only recently. There are, of course, many other methods. A summary of methods up to 2016 is available in the review of \citet{Ryan2016-yo}, since when, the field has progressed substantially, most notably through bounds on mutual information and methods that use these (covered in the next section). There are other more recent approaches which may in the future also advance \EIG estimation for geophysical applications (\eg \citet{Long2022-sn, Alexanderian2021-tx, Wu2020-dj, Goda2020-uo, Carlon2020-os, Beck2018-kg,Englezou2022-ed}). These are beyond our scope for this paper in which we have chosen those that currently currently appear most promising.

  \subsubsection{Mutual Information Bounds}

    By noting that the \EIG is equivalent to the mutual information between \m and \d, the developments in the estimation of mutual information can be used for Bayesian experimental design. Lower bounds on the mutual information (see \citet{Guo2021-hx} for a recent overview) are of particular interest for optimal design applications since they allow straightforward likelihood-free \EIG estimation (see subsection \ref{subsec:likelihood-free design}) and stochastic gradient design optimisation (see subsection \ref{subsec:stoch_gradient_opt}).

    The variational marginal and variational posterior methods presented above are examples of an upper \citep{Alemi2016-va} and a lower \citep{Barber2004-za} bound to the mutual information, respectively. Both share the problem of converging towards a biased estimate of the \EIG. The same is true for other \EIG estimators based on other mutual information lower bounds (\eg NWJ, InfoNCE, JSD \citep{Kleinegesse2021-aq}). However, those require a neural network with a single scalar output to be fitted instead of a variational mapping (which might be parametrised by a neural network) which is simpler to parametrise and carries less risk of misspecification \citep{Kleinegesse2020-ht}. 

    Mutual information bounds can also be constructed to be consistent, meaning they converge towards the true \EIG given a sufficient number of samples. An example of a consistent upper bound is the NMC method. By using importance or contrastive sampling, bounds based on variational mappings or neural networks can also be made consistent (\eg VNMC \citep{Foster2019-rx}, ACE \citep{Foster2019-ys}, FLO \citep{Guo2021-hx}). This consistency typically, but not necessarily (\eg FLO \citep{Guo2021-hx}), comes at the expense of more samples compared to unmodified methods but fewer samples compared to the NMC method \citep{Foster2019-rx}.

    The InfoNCE bound \citep{Van_den_Oord2018-qb}
    \begin{align}
      \operatorname{EIG}_\text{infoNCE}(\xi) = \frac{1}{K} \sum_{i=1}^K \log \frac{e^{T_{\boldsymbol{\phi}}\left(\mitbf{m}_i, \mitbf{d}_i\right)}}{\frac{1}{K} \sum_{j=1}^K e^{T_{\boldsymbol{\phi}} \left(\mitbf{m}_j, \mitbf{d}_i \right)}}
    \end{align}
    where $T_{\boldsymbol{\phi}}$ is a neural network with a single scalar output parametrised by $\boldsymbol{\phi}$ (optimised using SGD), is used to demonstrate \EIG estimation using MI lower bounds only parametrised by a neural network in section \ref{sec:applications}. While the FLO bound would be preferable due to its theoretical properties (consistency, low bias and variance), no readily available implementation was found at the time of writing.

%% file: EIG_optimisation.tex
The maximisation in Equation \eqref{eqn:eig_optimization} requires an optimisation algorithm to be chosen, and the choice determines how often the \EIG needs to be evaluated. Each evaluation can take a substantial amount of time to compute, so the choice of algorithm can dramatically change the overall computation required for experimental design.

\subsection{Global Optimisation}
  Ideally, the optimisation algorithm used to solve Equation \eqref{eqn:eig_optimization} should take all possible designs in $\Sigma$ into account and choose the one which results in the highest \EIG. This could be achieved most straightforwardly for a continuous design space by sampling $\Sigma$ in a sufficiently dense regular grid and calculating the \EIG for every grid point. However, this is only possible for low dimensional design spaces, because the number of \EIG evaluations scales as $(n_\text{grid})^D$, where $n_\text{grid}$ is the number of grid points per dimension and $D$ is the number of dimensions. To avoid this exponential scaling, several global optimisation algorithms such as the genetic algorithm \citep{Holland1992-ks} and simulated annealing \citep{Bohachevsky1986-qm} have been used to design geophysical problems (\eg \citet{Barth1992-ag, Barth1990-rd, Curtis1997-iq, Maurer1998-vd}). Their guarantee to always converge towards the global optimum given sufficient iterations comes at the cost of many \EIG evaluations, albeit fewer than is required by grid approach. This again makes these algorithms infeasible for all but small-scale design problems. Other popular global approaches for design optimisation in general are Bayesian optimisation \citep{Jones1998-bv, Foster2019-rx, Kleinegesse2018-zv} and MCMC methods \citep{Amzal2006-sb, Jones1998-bv}.

\subsection{Greedy Optimisation}

  The computational infeasibility of global optimisation has led to the use of greedy algorithms that make locally optimal choices but still lead to designs that substantially improve the \EIG compared to random designs. The most popular of those algorithms are \textit{sequential design optimisation} algorithms\footnote{Often also referred to as \textit{iterative design} in the statistical experimental design literature. Not to be confused with sequential design as introduced in section \ref{subsec:sequential_design}}, in which all but one dimension in the design space is fixed, and a one-dimensional optimisation is solved in the remaining dimension. Iterating this process, through all dimensions causes the design to converge to a (locally) optimal design.
  
  There are three main categories of sequential design optimisation. The most popular and computationally cheapest is sequential construction \citep{Curtis2004-uq, Stummer2004-lm} where the optimal design in a one-dimensional design space is selected first (\eg a first receiver is placed at an optimal location). This locally optimal choice is then fixed, and the best choice in a second one-dimensional design space (a second receiver location) is selected. This process can be iterated until the desired number of design dimensions (number of receivers) is reached. Alternatives to this approach include sequential destruction \citep{Curtis2004-uq, Curtis2004-jx} where design dimensions are sequentially removed from an exhaustive design space that includes all possible ways of acquiring data (\eg receivers at all possible locations) by decimating the design point with the smallest \EIG the advantage being that this method does consider all possible receiver configurations in iteration one (sequential construction never does). Also, the sequential exchange algorithm of \citet{Mitchell1974-gi}, is a hybrid, where one dimension (\eg a receiver location) is removed, then one is added in the same ways as above. All three sequential design optimisation algorithms are substantially cheaper than global optimisation, but in practice still usually result in designs that deliver an EIG that is close to the global optimum in practice \citep{Coles2011-dh, Guest2009-gi}. A detailed mathematical treatment of sequential design optimisation algorithms is given in \citet{Jagalur-Mohan2021-sc}. 

%% file: advanced_concepts.tex
We now briefly describe selected implementations of more advanced design concepts, using a set of examples that hint at possible useful developments in future. They also reinforce the notion that variational and other similar methods may be valuable for the design of experiments in certain types of geophysical applications.

\subsection{Likelihood-free experimental design} \label{subsec:likelihood-free design}
   
   So far, we have assumed that it is possible to evaluate \likelihood pointwise (meaning the likelihood function is explicit or otherwise directly computable). This assumption is typically valid in geophysics as the likelihood is often assumed to be a Gaussian distribution around a mean predicted by the forward model. However, under certain applications there exists inherent randomness in the forward function, in which case this is not possible.

   For model space methods this is of no consequence since the likelihood is only used to generate data samples and so its value need never be evaluated explicit. Data space methods, on the other hand, rely on explicit evaluations of \likelihood. If an average over nuisance variables can be used to model the intractable likelihood, the NMC method can be used in an extended form \citep{Feng2019-wp}, but in general this will lead to high computational cost in practical problems, because in this extended form, the inner loop samples can not be reused. Alternative methods include the use of a variational approximation of the likelihood (\eg \citet{Foster2019-rx, Cheng2020-ih}), and several everal works focus exclusively on experimental design algorithms for likelihood-free experimental design \citep{Kleinegesse2018-zv, Kleinegesse2020-ht, Kleinegesse2021-aq, Hainy2014-sv, Hainy2016-ck, Hainy2018-zo}.

\subsection{Designing Experiments for Interrogation Problems}
   
   \begin{figure}
      \centering
      \includegraphics[width=0.5\textwidth]{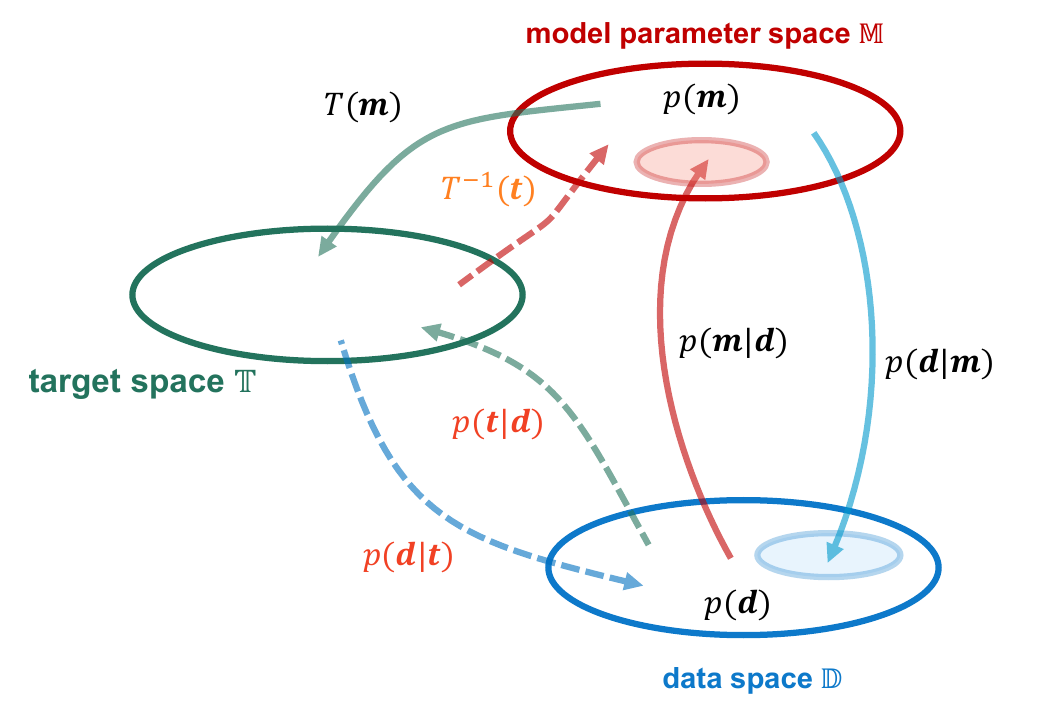}
      \caption{Schematic overview of model, data and target space and the (probabilistic) functions mapping between them.
      }\label{fig:interrogation_bed}
   \end{figure}

   The objective of a scientific investigation is typically to answer a specific set of research questions. For experimental design problems, we then wish to maximise the information in target space $\mathbb{T}$ rather than in the posterior distribution over model parameters. The answers usually depend on multiple model parameter values \m, and as shown in Fig. \ref{fig:interrogation_bed}, a target function $T(\mathbf{m}|Q)$ is defined that maps the values into a target space  $\mathbb{T}$ where question $Q$ can be answered \citep{Arnold2018-wx}.

   However, incorporating a target space poses challenges for data space experimental design methods since the likelihood $p(\mathbf{d} \, | \, \mathbf{t}) = \int_{\mathbb{M}} p(\mathbf{d} \, | \, \mathbf{t}, \mathbf{m}) \, p(\mathbf{m})$ is typically not available directly, because it depends on model parameter values. One approach to estimate the likelihood involves conditionally sampling models from the prior distribution that map to a specific point in the target space $p(\mathbf{m} \, | \, \mathbf{t})$, and treating the model parameters as nuisance variables \citep{Feng2019-wp}. For this approach, the inverse function $T^{-1}$ must be available, which is only the case in specific scenarios. Even if $T^{-1}$ can be approximated, it would involve similar challenges as previously discussed for variational methods. 
   
   By contrast, model space techniques enable straightforward likelihood-free experimental design (see section \ref{subsec:likelihood-free design}), and allow to design of interrogation experiments without requiring $T^{-1}$. They can therefore be applied for any general interrogation problem, for which $T$ or $T^{-1}$ is computable. Only if the mapping  $T$ is linear is it possible to use linear (bayesian) experimental design methods \citep{Curtis1999-mg,Wu2021-ye,Attia2018-gz}

\subsection{Stochastic Gradient EIG Optimization} \label{subsec:stoch_gradient_opt}

   Section \ref*{sec:EIG_opt} presents methods that solve Equation \eqref{eqn:eig_optimization} using search algorithms where \EIG calculation and optimisation are carried out in separate operations. This two-step approach is a standard procedure, but it becomes increasingly difficult for higher numbers of design dimensions. Using the variational posterior method, a one-step design procedure can be constructed in which the parameters of the variational family and the design vector are optimised simultaneously \citep{Foster2019-ys} using stochastic gradient descent. 
   
   SGD is a widely used optimisation algorithm which allows optimisations to scale to substantially higher dimensions. The only restriction for geophysical problems in practice is the need to provide the gradients of \likelihood with respect to the design. This limitation can be challenging for some problems but is readily available for problems that are solvable analytically (amplitude versus offset studies \citep{Van_Den_Berg2003-bn,Van_Den_Berg2005-lq}) or when the forward solver can be expressed in a backwards differentiable form (\cite{Smith2021-gk, Richardson2022-xl, Ren2020-nw}). Generally, any lower bound on the EIG (see \citet{Kleinegesse2021-aq, Foster2019-ys} for examples) can be straightforwardly maximised using SGD for design optimisation if the gradients of  \likelihood with respect to the design parameters are available or can be approximated. It is more challenging to apply upper bounds such as the NMC or variational marginal method in this one-step approach \citep{Foster2019-ys, Goda2020-uo, Huan2014-vy}.

\subsection{Sequential Design} \label{subsec:sequential_design}

   The previous sections assumed that only a single experiment was performed. While this is often true in geophysical studies, it neglects a significant advantage of Bayesian inverse theory: the possibility to iteratively update knowledge by taking the posterior information from a previous experiment as the prior for a following experiment. So-called sequential experimental design repeats this process for as many iterations as desired.
   
   This approach requires the repetition of a computationally expensive design process between experiments. Other techniques must be used if, after collecting data, the new design is needed quickly and without a significant computational expense (for example, in the case of an autonomous robotic system iteratively acquiring data, deciding which data to acquire next, then doing so). Deep adaptive design \citep{Foster2021-lg,Ivanova2021-ix} (similar work done by \citet{Pacheco2019-gy,Shen2021-jl,Shen2023-ka,Blau2022-ux}) trains a \textit{design network} which takes the results of previous experiments as inputs, account to produce a new optimal design in a single (cheap) forward pass of a neural network. While this form of sequential OED is a novel concept (especially in geophysics), there could be future applications in remote areas such as the deep ocean or on other planets where the use of autonomous measuring devices is necessary.

%% file: previous_work.tex
While linear and linearised OED has been used and studied extensivley for geophysical applications, OED for fully non-linear forward models is still not widely used. The first step in this direction was to use the number of modes in the misfit of linearised OED as a design criterion \citep{Curtis1999-id, Curtis2004-aw}. This method alleviates some problems of linearised designs.

The \EIG was first used as a criterion by \citet{Van_Den_Berg2003-bn, Van_Den_Berg2005-lq}. By adopting the maximum entropy sampling method they used the entropy of the evidence as a proxy for the \EIG. This approach has subsequently been refined to a sequential optimisation \citep{Guest2009-gi} and applied to design reflection seismic amplitude-versus-offset experiments with complex subsurface prior information \citep{Guest2010-zu}. The NMC formulation was first used in geophysics by \citet{Coles2012-kp} and has recently been applied by \citet{Qiang2022-ow} and combined with physics-informed neural networks by \citet{Wu2022-oz}. 

The Laplace Method \citep{Tierney1986-zg} allows the \EIG to be calculated in the model space using the Hessian matrix of the forward model (second order derivatives with respect to the model parameters), under the assumption that the posterior pdf is multivariate Gaussian \citep{Long2013-uj} (with extensions allowing for multimodality \citep{Long2022-sn}). It was used by \citet{Long2015-wl} for the optimal design of a full-waveform moment tensor inversion. The use of the determinants of Hessian matrices makes this method similar to linearised Bayesian experimental design studies.

The computationally efficient $D_n$ method first used by \citet{Coles2011-ks} has subsequently been applied and derived using alternative approaches \citep{Rawlinson2012-cs, Coles2013-vw, Bloem2020-gp}. In addition to the \EIG, other studies on non-linear design have used bifocal measures \citep{Winterfors2008-ay, Winterfors2012-uc}, or problem-specific measures which are not a function of the posterior pdf \citep{Lopez-Comino2017-pj, De_Landro2020-zf, Ferrolino2020-jl, Fichtner2022-wi, Dasgupta2021-in}.

More detailed reviews of design methods and applications in geophysics are given in \citet{Curtis2004-jx, Curtis2004-aw} and \citet{Maurer2010-jt}. For a general review of design algorithms, the reader is referred to \citet{Ryan2016-yo}.

%% file: applications_srcloc.tex
We now demonstrate the algorithms introduced in detail above, and explore their relative merits in two common geophysical problems. Our aim here is to be educational rather than to provide a comprehensive study of these optimal design problems. 

\subsection{Seismic Source Location}
  
\begin{figure*}
  \centering
  \includegraphics[width=\textwidth]{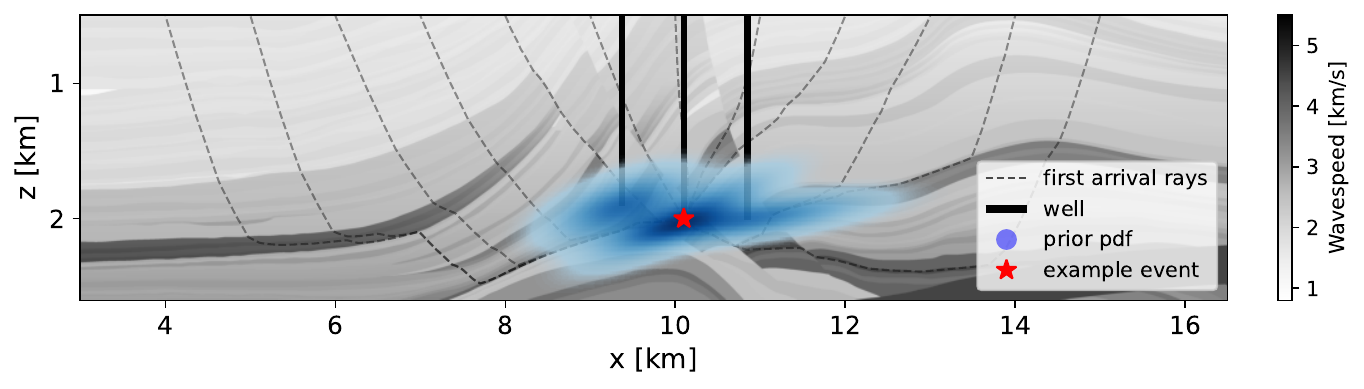}
  \caption{
    Seismic source location problem with three wells (thick black lines), prior pdf comprising a sum of three spatially correlated Gaussians (blue shading, darker being more probable) and seismic first arrival rays (thin black lines) originating from one of the prior samples (red star) and terminating at regularly spaced points on the ground surface.}\label{fig:setup_gmm}
\end{figure*}

In a seismic source location problem, the aim is to determine the location of a seismic event such as an earthquake using the first arrival times of P waves, S waves or other seismic phases. To achieve this, the subsurface structure needs to be known at least approximately.

The setup used in this example (depicted in Fig. \ref{fig:setup_gmm}) employs the elastic extension of the two-dimensional Marmousi model \citep{Martin2006-jf} a suitably complex subsurface structure. Three wells with depths of around 2~km are placed near the middle of the structure, and are assumed to be involved in some intervention in the subsurface which induces seismicity near their terminations. As the locations of the wells are known, the model parameter prior pdf is constrained in space, and we represent our assumed prior information as a sum of three spatially correlated Gaussians (Fig. \ref{fig:setup_gmm}).

The strong prior information combined with the 2D nature of this synthetic example, allows a low number of receivers to achieve a good posterior estimate of the true location. Under the assumption of a constant $v_p/v_s$ ratio and using arrival time differences $t_ \mathrm{diff}$ between P and S waves, the source time can be excluded from the source location problem similarly to \citet{Bloem2020-gp}, and hence can be excluded from the design process. Therefore, the model parameter space consists of the set of two-dimensional vectors of horizontal and vertical locations. 

The fast marching method \citep{Sethian1996-pd} using the openly available implementation of \citet{White2020-nh} was used to calculate the seismic wave arrival times. \citet{Bloem2020-gp} used the same travel time inversion to test different linear and nonlinear experimental design algorithms, albeit with other slowness models and prior pdfs. The likelihood is modelled using a Gaussian distribution with a mean corresponding to the calculated travel time and a standard deviation of 0.02~s as a baseline for the measurement uncertainty. 

The optimal design problem is to find receivers placed on the seabed (z=0.5~km) at horizontal offsets from 3.1-16.4~km, which gives the highest \EIG. The \EIG of 200 possible single-receiver locations was calculated along the seabed using the various methods described in section \ref{sec:EIG_est}. The sequential construction method was used to optimise multi-receiver designs because of its computational efficiency and the possibility of visualising the design process steps as receivers are added.

\begin{figure*}
  \centering
    \includegraphics[width=1.0\textwidth]{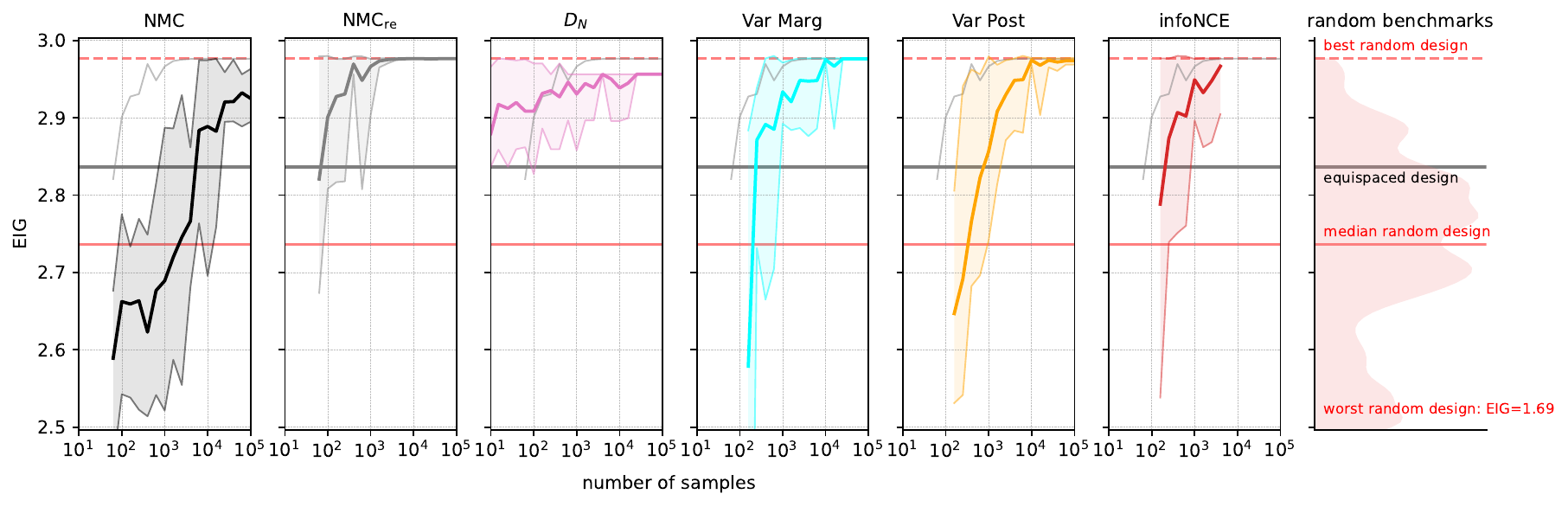}
    \caption{
      Benchmark results from different \EIG estimation and design methods, showing the \EIG for a two-receiver network for seismic source location as a function of the number of samples for which the forward function is evaluated. The solid line indicates the mean of 10 independent runs, while the shaded area indicates the respective minimum and maximum values of those runs. The mean curve of the $\text{NMC}_\text{re}$ results are shown in every panel to serve as a benchmark for comparison. On the right, a smoothed histogram of the \EIG for 1000 randomly selected designs and for a heuristic design with receivers at 6.4 and 13.1~km is shown to put the results into context. Details on the setup and methods are in the main text.
    }\label{fig:srcloc_benchmark_eig_2_kde}
  \end{figure*}

  In most geophysical applications, the generation of data corresponding to model parameter samples is the main computational cost. We therefore compare the performance of the different methods for a two-receiver network as a function of the number of samples used (see Fig. \ref{fig:srcloc_benchmark_eig_2_kde}). The cost of evaluating the different estimators using those samples is compared thereafter.

  The variational marginal method has been implemented using a Gaussian Mixture Model (GMM) with 10 Gaussians, each with a full covariance matrix. For the variational posterior method, the variational family is the output of a Mixture Density Network (MDN) with a three-layer neural network consisting of 60 nodes in each layer, defining 20 Gaussians with full covariance matrices as output. The InfoNCE method has been implemented using a three-layer neural network consisting of 40 nodes in each layer.
  
  The NMC method has been implemented to use new samples for each step of the outer loop or to reuse the $M$ samples of the inner loop of Equation \ref{eqn:NMC} (the latter is referred to as $\text{NMC}_\text{re}$). This results in a cost of $N \times M$ or $N+M$ samples, respectively.
  
  To put the results into context, we calculated the \EIG for 1000 random designs, and for a design with receivers at 6.4 and 13.1~km, which is a proxy for a heuristically designed network. All methods converge to a \EIG value that outperforms almost all of the random designs as well as the heuristic design. Due to the restrictive Gaussian assumption, the $D_N$ method has a slight negative offset. At the cost of performing slightly worse compared to the other methods (apart from NMC), the $D_N$ method design substantially outperforms the median and equispaced design with as few as ten samples. At this low number, all other methods are practically inapplicable. Both variational methods perform similarly to the $\text{NMC}_\text{re}$ method, converging towards the maximum \EIG when using more than around $1\tento{4}$ samples. 
  
  Tests using a four-receiver design lead to similar conclusions, with the exception that both the variational posterior and the InfoNCE method perform better and are comparable to the $\text{NMC}_\text{re}$ method, demonstrating the beneficial scaling of those methods with data dimensionality. For three and four receivers, all methods converge to designs with slightly lower \EIG compared to the best random design, which is an effect of using the sequential construction method which can converge to local maxima.
  
  \begin{figure}
    \centering
    \includegraphics[width=0.5\textwidth]{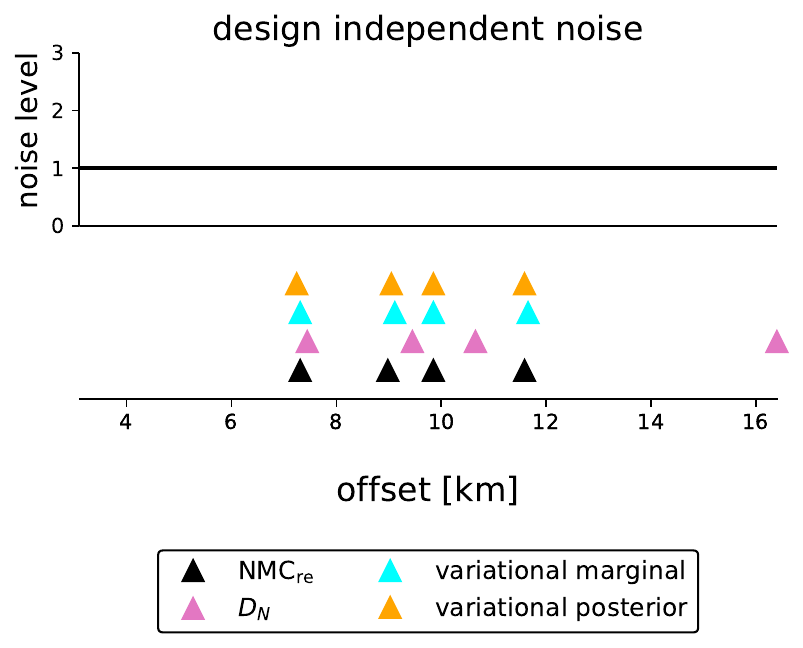}
    \caption{
      Optimised four-receiver networks using different \EIG estimation methods and the sequential construction method for a constant noise level. The upper part shows the noise level as a function of receiver position in multiples of the base (constant) noise level which has standard deviation 0.02~s.
    }\label{fig:srcloc_noise_comparison_results}
  \end{figure}

  The final designs for a four-receiver network are shown in Fig. \ref{fig:srcloc_noise_comparison_results}. For all but the $D_N$ method, the resulting network is nearly identical, and the deviation of the $D_N$-derived network leads to the slight negative offset in \EIG discussed above.

  \begin{figure}
    \centering
    \includegraphics[width=0.5\textwidth]{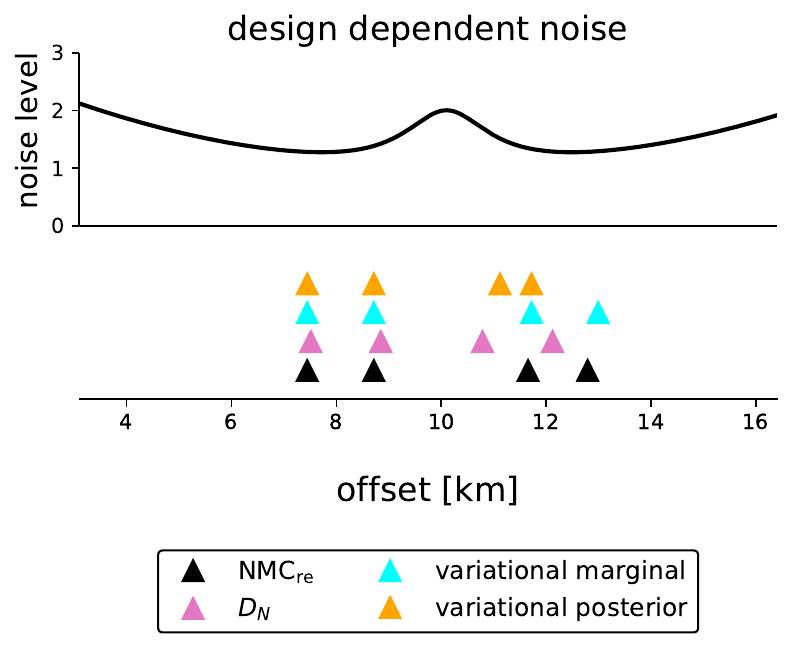}
    \caption{
      Optimised four-receiver networks using different \EIG estimation methods and the sequential construction method for a design-dependent noise level. The upper part shows the noise level as a function of receiver position in multiples of the base (constant) noise level which has standard deviation 0.02~s.
    }\label{fig:srcloc_noise_comparison_results_noise}
  \end{figure}

  For now, we have assumed a constant noise level across all receiver locations, or in other words we have assumed the noise to be independent of the design. This is rarely the case in real-world scenarios. Fig. \ref{fig:srcloc_noise_comparison_results_noise} shows the effect of design-dependent noise. Here, the experimental design process uses an artificial noise function that mimics the effects of geometrical spreading and of anthropogenic noise around the wells. The best designs found change substantially, moving receivers towards regions of low noise, leading to a higher agreement in the optimal design results from different methods. The design dependence of the likelihood also stabilised the design optimisation process, and indeed the result in this case might have been designed using intuition alone. This demonstrates that where intuition can be applied, the design methods herein conform to expectations. And of course, a quantitative approach is necessary for more complex noise models and realistic three-dimensional environments in which intuition fails.

  While the number of forward evaluations is often the bottleneck for the feasibility of experimental design algorithms, the cost of the \EIG estimator itself is not insignificant. This is especially important in problems where physical insights can be used to generate a large number of data samples. In the case of the presented source location example, this is possible by making use of the reciprocity of the eikonal equation and the full traveltime field generated by the fast marching method used to solve the eikonal equation: by treating receiver locations as eikonal source fields, data and model parameter pairs can be generated cheaply once the travel time field for a given receiver location has been calculated. After the precomputed travel times are stored for each possible receiver position, the main bottleneck in design optimisation is the cost of the \EIG estimator.

  \begin{figure}
    \centering
    \includegraphics[width=0.5\textwidth]{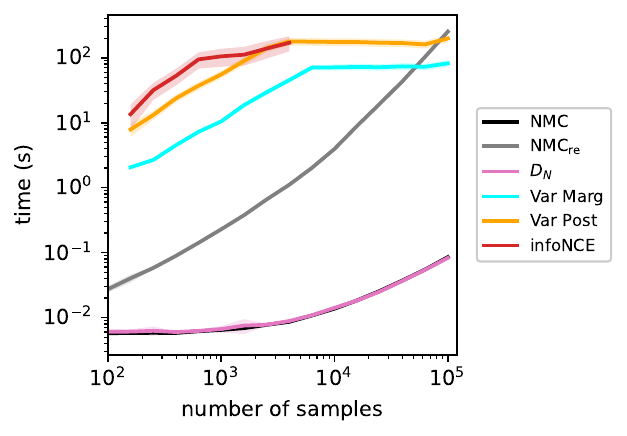}
    \caption{
      Benchmark of different \EIG estimation methods. Shows the time to calculate the \EIG for a two-receiver network for seismic source location, excluding the time to generate data samples. The solid line indicates the mean of 10 independent benchmark runs, while the shaded area indicates the respective minimum and maximum values. Details on the setup and methods are in the main text.
      }\label{fig:srcloc_benchmark_times_2}
  \end{figure}

  The cost of the \EIG calculation, excluding the cost of the forward evaluations, is shown in Fig. \ref{fig:srcloc_benchmark_times_2}. All six methods become increasingly more expensive the more samples are used. The $D_N$ and NMC methods are substantially cheaper than the other methods. Both variational and MI lower bound methods have a relativley large overhead since they require neural networks to be trained to represent the variational approximators. Since the training of the variational estimators is the main cost in using them for \EIG estimation, they scale very well for a high number of samples if the maximum number of stochastic gradient descent steps is capped (here 10000 for the variational methods and 1000 for the InfoNCE method). The point where this threshold is reached can be seen as the sharp change in slope for those methods in Fig. \ref{fig:srcloc_benchmark_times_2}. The difference in computation time between the NMC with and without reused samples is twofold. First if $N_T$ is the total number of samples used, the standard NMC method requires $N_T$ likelihood evaluations, while the NMC with reused samples requires $0.25 \times {N_T}^2$. This quadratic scaling is evident in Fig. \ref{fig:srcloc_benchmark_times_2}. A second difference is that a numerically slightly less efficient method of computing the NMC with reused samples was used since otherwise the memory usage for the computation of the \EIG for $1\tento{5}$ samples would be more than 20GB for storing the necessary likelihood values alone.

  \subsubsection{Drawbacks of the $D_N$ method}
  
    \begin{figure*}
      \centering
      \includegraphics[width=1.0\textwidth]{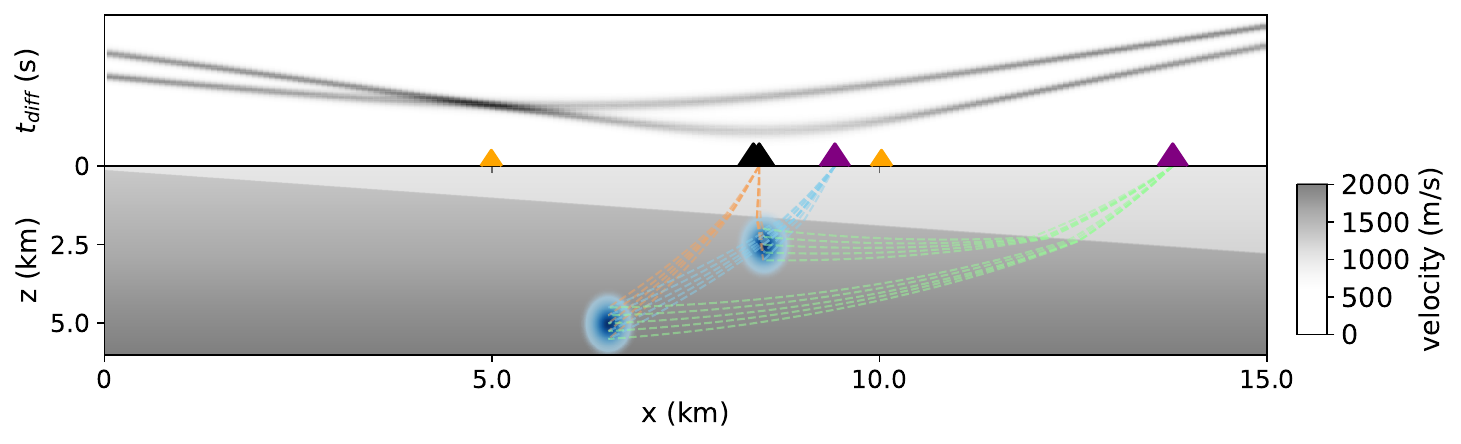}
      \caption{
        Setup for demonstrating shortcomings of the $D_N$ method. The lower part shows the subsurface P-wave velocity model, including the prior pdf for event locations (blue contours). The dotted lines give examples of first arrival rays originating from ten representative prior locations to three receiver locations. The optimal design calculated using the NMC method (black triangles) and the $D_N$ method (purple triangles) are shown, as well as a heuristic equispaced design (orange triangles). The top part shows a histogram of travel times generated by forward modelling samples from the prior pdf for each possible receiver location. Darker grey tones indicate that more samples produce the same travel time.}\label{fig:break_dn_setup_evidence}
    \end{figure*}

    The above results established the $D_N$ method as a cheap and robust method that can produce near-optimal designs. This is also confirmed in the later section \ref{subsec:AVO} for an AVO design problem. In both cases, the assumption of a Gaussian form for the evidence only leads to a slight negative offset, yet we investigate conditions under which this assumption nevertheless lead to the $D_N$ method to produce a far from optimal design. We set up a scenario (see bottom of Fig. \ref{fig:break_dn_setup_evidence}) with two separate areas of possible seismicity, with an inclined low-velocity layer above. For example, the cause of the seismicity might be related to drilling or injection for geothermal power production (\eg \citet{Maurer2020-yb}).

    \begin{figure}
      \centering
      \includegraphics[width=0.4\textwidth]{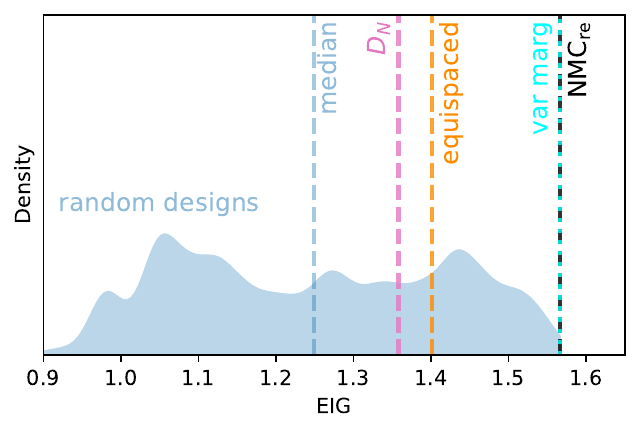}
      \caption{
         Comparison of the \EIG values for two-receiver networks derived using the NMC (reused inner samples), the variational marginal, and the $D_N$ methods as well as an equispaced heuristic design (see Fig. \ref{fig:break_dn_setup_evidence}). The blue smoothed histogram represents the \EIG values of 1000 randomly selected designs.
        }\label{fig:break_dn_eig_comparison}
    \end{figure}

    In this scenario, the two receiver-network designed using the $D_N$ method  is quite different (Fig. \ref{fig:break_dn_setup_evidence}) and performs substantially worse (Fig. \ref{fig:break_dn_eig_comparison}) than one designed using either the NMC (reused $M$) or the variational marginal method (using a mixture of Gaussians). While the $D_N$ design still performs better than the median of 1000 random designs, around a third of the randomly selected designs perform better, and it performs worse than a standard design in which the receivers are equispaced at 5000 and 10000m. The distribution of travel times as a function of horizontal receiver position illustrates why the $D_N$ method performs so poorly in this scenario (top of Fig. \ref{fig:break_dn_setup_evidence}). As long as the two bands of travel times due to the two areas of seismicity overlap with extremely low probability, their information content is independent of the distance between them. However, if the two bands are further apart, the standard deviation of the evidence increases, and therefore the information content of the Gaussian used to approximate the evidence in the $D_N$ method decreases. Due to the nature of this scenario, the distance between the two bands varies substantially across the different receiver placements. Therefore, the $D_N$ method is, to first order, influenced by the distance between the travel-time bands and not their respective spread, whereas the latter property governs the information content of the evidence. In applications where substantial multimodality may occur \textit{a priori} in data space, this deficiency trades off with the computational efficiency of the method. 

  \subsubsection{Source Location Interrogation}\label{subsec:srcloc_interrogation}

    As mentioned in section \ref{sec:var_post}, solving the \EIG estimation in model parameter space allows one to design interrogation problems \citep{Arnold2018-wx}. Instead of maximising the information of \posterior, the goal in such problems is to maximise the information in a target space $\mathbb{T}$, which is used to answer a specific question or set of questions $Q$. For this, a target function $T(\mitbf{m} \, | \, Q)$, which maps samples from $\mathbb{M}$ to $\mathbb{T}$, needs to be defined.

    We apply this concept to search for the optimal receiver placements to constrain only the epicentre or only the depth of the seismic source, respectively. In this case, $T$ selects and returns one of the two coordinates, making the target space one-dimensional ($\mathbb{R}^1$). 
    
    \begin{figure*}
      \centering
      \includegraphics[width=1.0\textwidth]{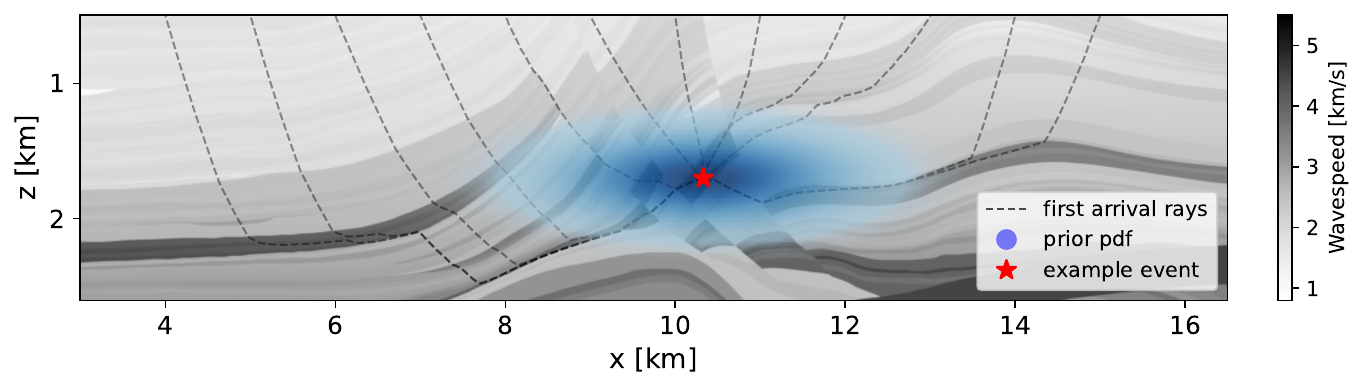}
      \caption{Seismic source location problem setup with prior samples (blue contours) and seismic rays (thin black lines) originating from one of the prior locations (red star).}\label{fig:setup_basic}
    \end{figure*}
    
    A slightly simpler setup than the one above is used to demonstrate the interrogation design process; see Figure (\ref{fig:setup_basic}). The same constant noise level as in the previous test was used. The only thing that changed is the prior, now a single multivariate Gaussian distribution, placed in the relatively shallow subsurface.

    \begin{figure}
      \centering
      \includegraphics[width=0.5\textwidth]{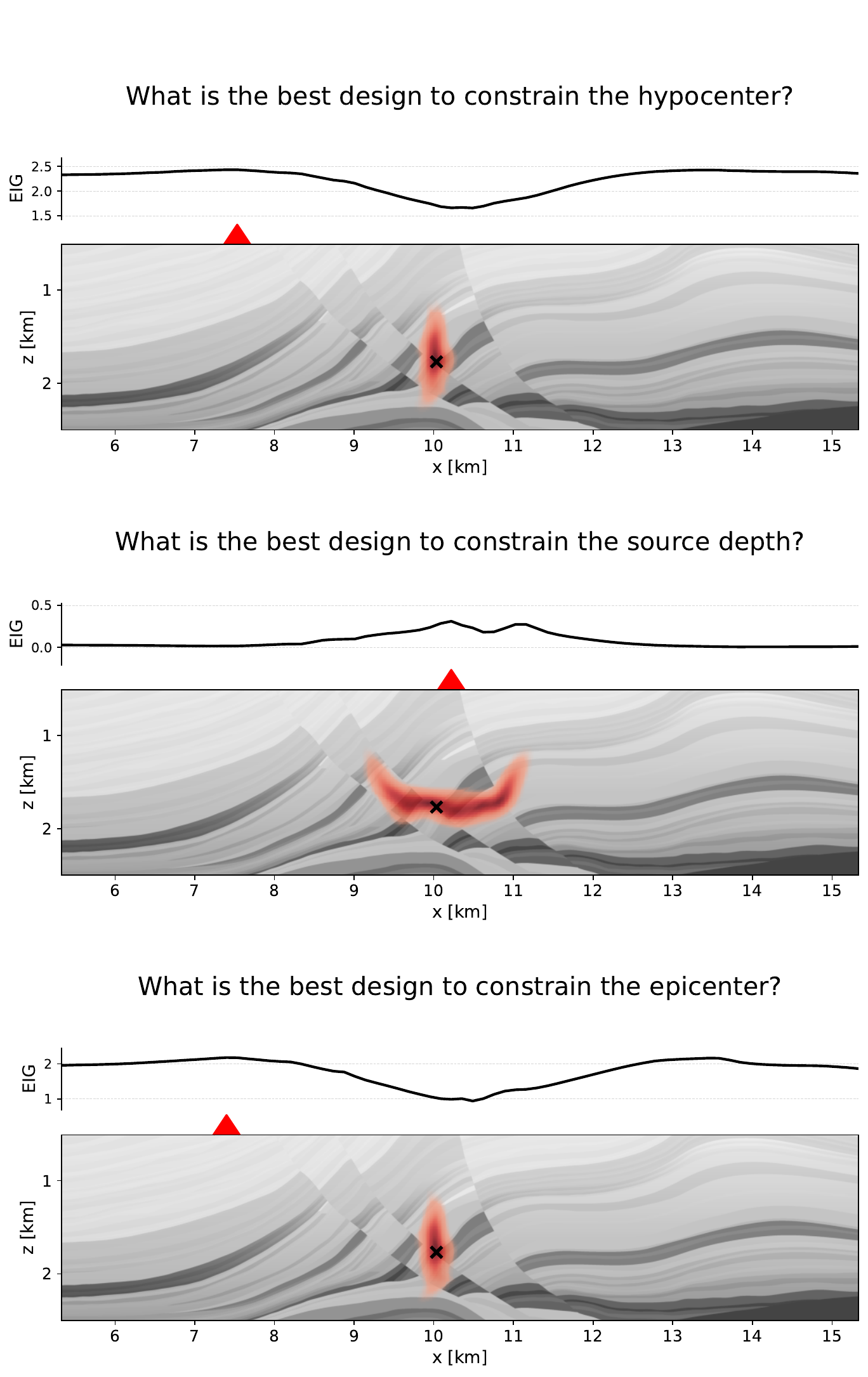}
      \caption{
        Comparison of optimal one-receiver networks for different interrogation goals. A high \EIG in the top graph of each panel indicates good positions for answering the specific questions given in the panel title. The bottom graph of each scenario illustrates the geophysical setting and shows an example model parameter posterior pdf. The black cross indicates the true event location used to generate the datum in each example.
      }\label{fig:interrogation_summary_1rec}
    \end{figure}

    Figure (\ref{fig:interrogation_summary_1rec}) compares example posterior probability distributions computed for three designs, each optimised for a different interrogation question. The \EIG curves show that both the hypocenter and the epicentre design problems favour nearly identical designs with a low \EIG over the source area and higher \EIG at a greater distance from the expected sources. If the vertical location (source depth) is to be constrained independently, the resulting \EIG curve and optimal receiver position are entirely different: Here receiver positions over the source area are preferred, with positions at larger distances providing nearly no information. The resulting posterior probability functions also clearly show the effect of different receiver positions. The area of high probability is aligned vertically when we focus on hypo- or epicentre, while it is aligned almost horizontally when we seek the source depth. An information tradeoff introduced by a focus on different questions is also evident since the posterior pdf for the source depth is more spread out and therefore is less informative than the hypocenter localisation design, which corresponds to classical experimental design.

%% file: applications_avo.tex
\subsection{Amplitude versus Offset}\label{subsec:AVO}
  
  \begin{figure}
    \centering
    \includegraphics[width=0.4\textwidth]{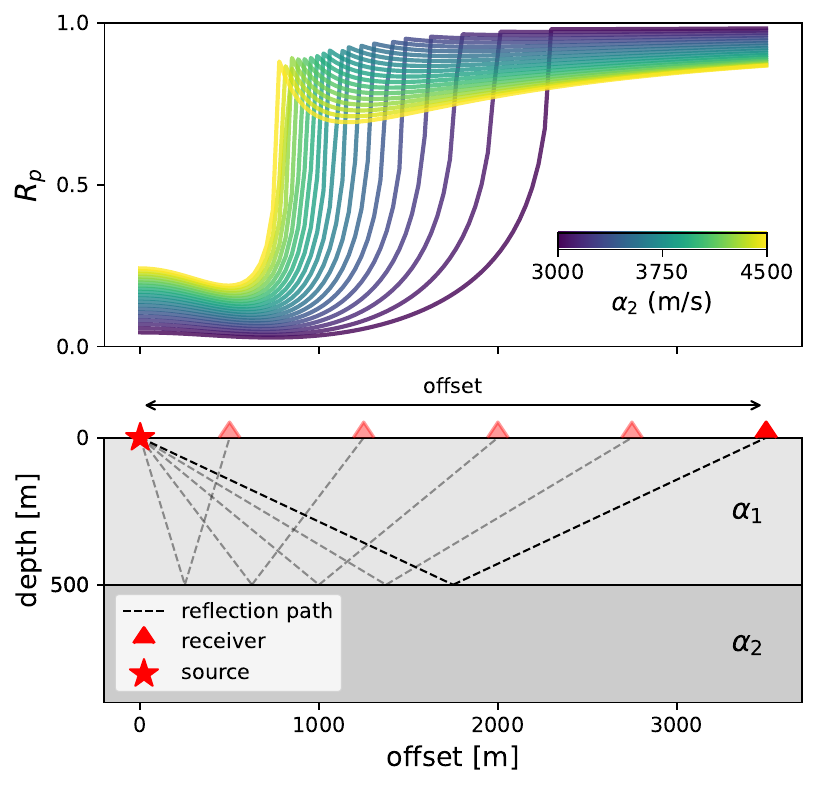}
    \caption{
      Schematic illustration of the amplitude versus offset (AVO) design problem setup. The top figure shows the change in reflection coefficient $R_p$ (dimensionless) as a function of offset for different P-wave velocities in the lower layer, indicated by the different colours.
    }\label{fig:avo_setup}
  \end{figure}

  A well-studied problem in non-linear geophysical optimal design is amplitude versus offset (AVO) inversion for seismic velocity contrasts \citep{Van_Den_Berg2003-bn,Van_Den_Berg2005-lq, Guest2009-gi, Guest2010-zu, Guest2011-eu}. We use this example to compare methods and discuss contrasts with results in the source location problem, and to illustrate stochastic 1-step design optimisation and a more realistic interrogation problem.
  
  The objective of AVO is to determine the seismic properties of a buried layer by observing the change in seismic amplitude as a function of offset from the source. Fig. \ref{fig:avo_setup} depicts the setup which is identical to the one in \citet{Guest2009-gi}. A Gaussian prior pdf with a mean of 3750~m/s and a standard deviation of 300~m/s is assigned to the P-wave velocity $\alpha_2$ in the layer of interest. Further, we assume a so-called Poisson medium in which $\beta=c\alpha$, where $c=1/\sqrt{3}$, no density contrast between the layers, and assign a P-wave velocity $\alpha_1$ of  2750~m/s and thickness of 500~m to the upper layer. The top of Fig. \ref{fig:avo_setup} shows the resulting reflection coefficients for a range of offsets and values for $\alpha_2$, indicating the non-linear nature of the problem, especially around the critical angle.

  \begin{figure*}
    \centering
      \includegraphics[width=1.0\textwidth]{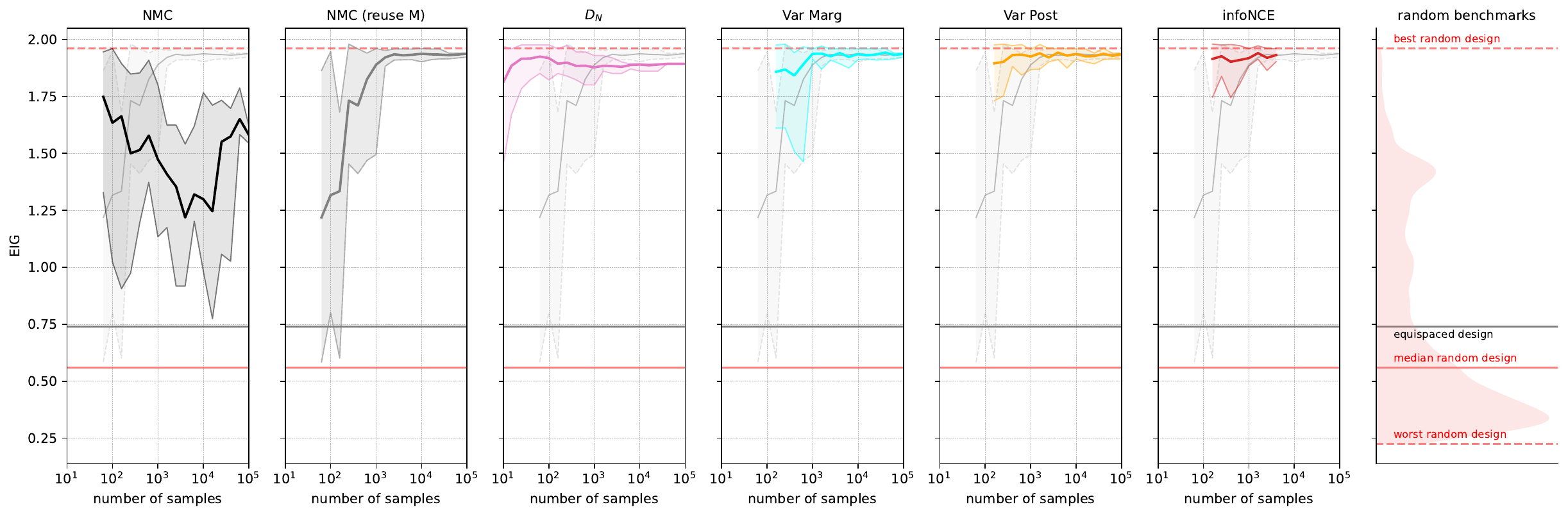}
      \caption{
        Benchmark of different \EIG estimation methods. Shows the \EIG for a two-receiver network for an AVO design problem. The solid line indicates the mean of 10 independent benchmark runs, while the shaded area indicates the respective minimum and maximum values of those runs. The $\text{NMC}_\text{re}$ results are shown in every panel to serve as a benchmark for comparison. On the right, a smoothed histogram of the \EIG for 1000 randomly selected designs and the EIG of a heuristic design is shown to put the results into context. Details on the setup and methods are in the main text.
        Benchmark results from different \EIG estimation and design methods, showing  the \EIG for a two-receiver network for an AVO design problem as a function of the number of samples for which the forward function is evaluated. The solid line indicates the mean of 10 independent runs, while the shaded area indicates the respective minimum and maximum values of those runs. The mean curve of the $\text{NMC}_\text{re}$ results are shown in every panel to serve as a benchmark for comparison. On the right, a smoothed histogram of the \EIG for 1000 randomly selected designs and for a heuristic design with receivers at 0.5~km and 1.5~km is shown to put the results into context. Details on the setup and methods are in the main text.
      }\label{fig:avo_benchmark_eig_2_kde}
    \end{figure*}

    As for the seismic source location example, the different methods presented in section \ref{sec:EIG_est} are compared as a function of the number of forward samples used. The only changes compared to the seismic source location benchmarks are that less expressive variational families are used for the variational posterior method (MDN with three layers consisting of 30 nodes in each layer, defining 10 Gaussians) and the InfoNCE method (neural network with three layers consisting of 20 nodes in each layer).

    All variational methods perform substantially better if the $\text{NMC}_\text{re}$ method is taken as a baseline. Especially the methods based on \EIG lower bounds (variational posterior and InfoNCE) perform well in this scenario. If inner loop samples are not reused, the NMC method performs poorly, showing no sign of convergence even for $1\tento{5}$ total samples. This is likely due to the very low sampling density in regions with near vertical $R_p$ curves leading to near zero probabilities. Since even for $1\tento{5}$ total samples the inner loop contains only 46 samples the NMC \EIG estimate has a high variance. For $\text{NMC}_\text{re}$ around $1\tento{3}$ inner loop samples are necessary to converge. If the NMC method without reused samples behaves similar in the order of $1\tento{9}$ samples would be necessary for it to converge.

    All variational methods perform substantially better if the $\text{NMC}_\text{re}$ method is taken as a baseline. Especially the methods based on \EIG lower bounds (variational posterior and InfoNCE) perform well in this scenario. If inner loop samples are not reused, the NMC method performs poorly, showing no sign of convergence even for $1\tento{5}$ total samples. This is likely due to the very low sampling density in regions with near vertical $R_p$ curves leading to near zero probabilities. Since even for $1\tento{5}$ total samples the inner loop contains only 46 samples the NMC \EIG estimate has a high variance. For $\text{NMC}_\text{re}$ around $1\tento{3}$ inner loop samples are necessary to converge. If the NMC method without reused samples behaves similar in the order of 
    All variational methods perform substantially better if the $\text{NMC}_\text{re}$ method is taken as a baseline. Especially the methods based on \EIG lower bounds (variational posterior and InfoNCE) perform well in this scenario. If inner loop samples are not reused, the NMC method performs poorly, showing no sign of convergence even for $1\tento{5}$ total samples. This is likely due to the very low sampling density in regions with near vertical $R_p$ curves leading to near zero probabilities. Even for $1\tento{5}$ total samples the inner loop of the NMC method contains only 46 samples which leads to a high variance of the \EIG estimate. Around $1\tento{9}$ total samples would be necessary for the NMC method to have the same number of inner loop samples as the $\text{NMC}_\text{re}$ method at the point where it first convergences to a stable design ($1\tento{3}$ samples). The \EIG value to which all methods converge is slightly lower than the best design of 1000 random trials due to the use of sequential construction for optimisation. However, for a four-receiver design, the optimal design of all but the NMC methods outperform the best random design by a substantial margin (around 10\%). The value of experimental design is obvious, considering how much worse the average random designs and the heuristic design perform.
    
  \subsubsection{One step \EIG optimisation}
    Using variational lower bounds, we use the AVO experimental design problem to illustrate the feasibility of stochastic gradient descent design optimisation for geophysical problems. We use the variational posterior method and its gradients with respect to receiver positions to optimise the design $\xi$ while simultaneously fitting parameters $\boldsymbol{\phi}$ describing the variational approximator $q_p(\mitbf{m} \, | \, \mitbf{d}, \xi, \boldsymbol{\phi})$. This approach is the BA-method introduced by \citet{Barber2004-za} and applied to experimental design first by \citet{Foster2019-ys}. The variational family is the output of a MDN with a three-layer neural network of 40 nodes in each layer defining ten Gaussians as output. The design $\xi$ is passed as an additional input to potentially strengthen the extrapolation capabilities of the variational mapping as proposed by \citet{Kleinegesse2021-aq} for a neural-network-only based \EIG lower bound. $1 \tento{5}$ SGD steps with a batch size of one and the optimiser ADAM \citep{Kingma2014-gr} were used to find values of $\xi$ and $\boldsymbol{\phi}$ which maximise the \EIG. 

    \begin{figure*}
      \centering
      \includegraphics[width=1.0\textwidth]{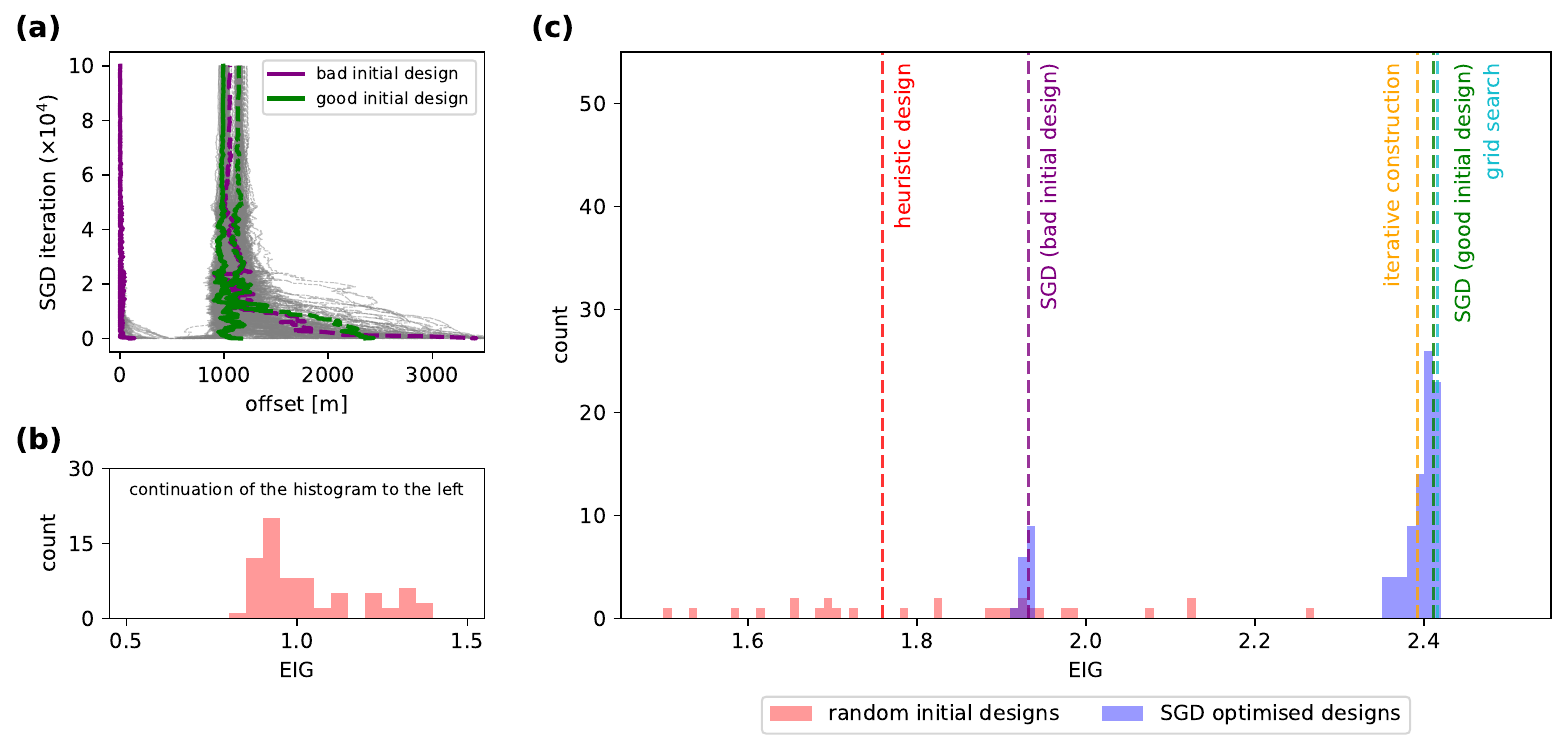}
      \caption{
        Summary of the results for the design of a two-receiver network for the AVO problem using stochastic gradient descent (SGD) design optimisation (c). \EIG values of designs derived using the sequential construction method, a grid search, and a heuristic design are given for comparison. All \EIG values are calculated using the $\text{NMC}_\text{re}$ method. (a) Offsets of the two receivers during the SGD design optimisation, where solid lines indicate the receiver starting with the lower offset, and dashed lines the one starting with the higher one. Red and blue histograms correspond to the \EIG of starting and final designs, respectively. (b) Continuation of panel (c) towards the left with a lower resolution.
      }\label{fig:avo_onestep_2rec_designs_eig_benchmark}
    \end{figure*}

    As a first test, we search for an optimal two-receiver design which can be benchmarked against a grid search using the $\text{NMC}_\text{re}$ method ($N = 1 \tento{4}$, $M = 1 \tento{4}$). Since gradient descent algorithms are sensitive to local extrema, the initial design choice is important. We therefore ran 100 trials with random initial offsets (Fig. \ref{fig:avo_onestep_2rec_designs_eig_benchmark}) where the \EIG at the final design has been calculated using the $\text{NMC}_\text{re}$ method to make the results comparable. Most designs using SGD design optimisation converge to the maximum \EIG of the $200\times200$  grid search which is assumed to approximate the global maximum. The design calculated by sequential construction selecting locations from 200 offsets performs slightly worse than the grid search and most SGD-optimised designs. In less than $20\%$ of cases, one receiver gets stuck at the local minimum at zero offset. The $R_p$ versus offset graph in Fig. \ref{fig:avo_setup} clearly shows this local maximum since the spread in data increases towards lower offsets from around 650~m. Due to the boundary at zero offset and the region with small gradients up to offsets of around 650~m, it is hard for the SGD algorithm to escape this local maximum in \EIG. This behaviour can be seen in Fig. \ref{fig:avo_onestep_2rec_designs_eig_benchmark}(a), which shows the change in designs during the SGD optimisation process. In any case, the SGD-optimised designs outperform a heuristic design with equispaced receivers at 750~m, and 1250~m, since heuristically it is beneficial to place receivers between one and three times the depth of the interface \citep{Guest2009-gi}. Initial designs with receivers at [50~m, 3450~m] (good initial design) or at [1166~m, 2333~m] (bad initial design) show the effects of choosing a reasonable or unreasonable initial design choice, where the unreasonable one is stuck in a local maximum while the reasonable one converges towards the global maximum. 
    
    At the expense of potential convergence towards a local maximum, the SGD design optimisation requires only $1 \tento{5}$ forward evaluations compared to the $8 \tento{8}$ for the grid search and $8 \tento{6}$ evaluations for the sequential construction using the $\text{NMC}_\text{re}$ method -- resulting in a reduction of computational cost by a factor of 8000 and 80 respectively. If repeated computations are stored, the reduction in the number of forward evaluations drops to a factor of 20 in both cases, but savings would increase if a finer grid is used. The actual savings are hard to estimate since calculating the \EIG using the $\text{NMC}_\text{re}$ method still involves a large number of likelihood evaluations, even if the forward model evaluations are precomputed. At the same time, the SGD algorithm introduces the overhead of calculating the gradients of the mixture density network for each sample using automatic differentiation. The actual reduction depends on the computational cost of the forward model, the complexity of the variational family, and the number of samples necessary for getting a stable \EIG estimate.

    \begin{figure}
      \centering
      \includegraphics[width=0.5\textwidth]{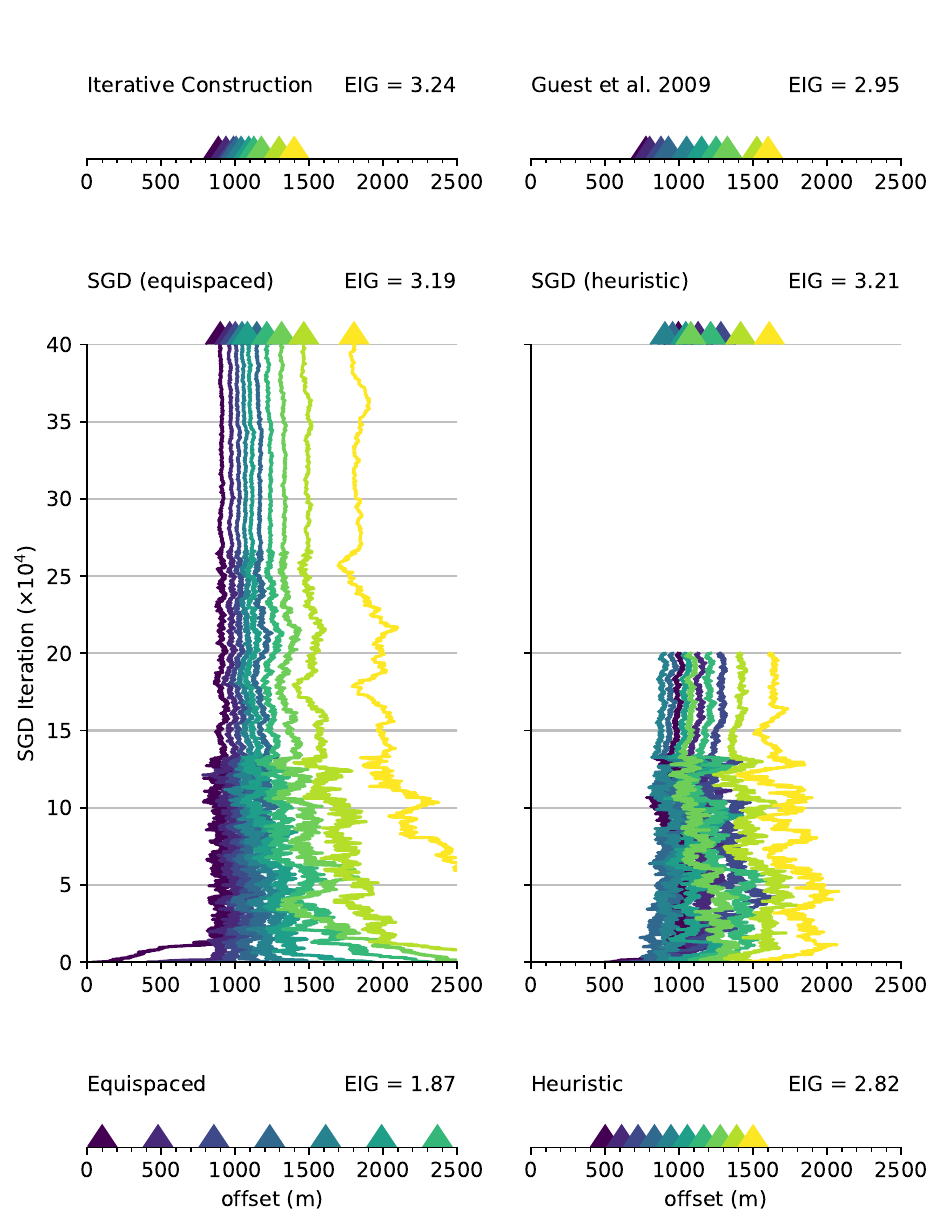}
      \caption{
        Designs (coloured triangles) and \EIG values for ten-receiver SGD design optimisation for the AVO design problem. If SGD iterations are shown on the y-axis, the design and \EIG corresponds to the final (uppermost) design. SDG (equispaced) refers to an equispaced design between 0 and 3500~m which is optimised using $1 \tento{5}$ SGD steps, while SDG (heuristic) refers to a heuristically good design with equispaced receivers between 1500 and 2500~m which is optimised using $5 \tento{4}$ SGD steps. The offset axis is limited to a range from 0 to 2500~m since the receivers for all but the equispaced design are concentrated in this area. 
      }\label{fig:avo_onestep_comparison}
    \end{figure}

      To display the beneficial scaling properties of SGD design optimisation, the AVO experimental design problem was repeated with ten receivers and compared to the results of \citet{Guest2009-gi}. We used an MDN with a three-layer neural network of 100 nodes in each layer defining 20 Gaussians as output. In this case, the SGD optimisation algorithm was substantially less likely to get stuck in local minima and converged to a consistent solution in nearly all test runs. The number of SGD steps was increased to $4\tento{5}$ to accommodate the larger number of receivers. However, as is clear from Fig. \ref{fig:avo_onestep_comparison}, the SGD optimised design is already very close to the final design after around $5\tento{4}$ iterations. 

      Two different initial conditions were tested as starting points for the SGD optimisation. First, we used ten equispaced receivers between 50 and 3450~m, which introduces little prior knowledge and is only subject to the constraint that receivers should lie between an offset of 0 and 3500~m. Second, the initial design are equispaced receivers between 500 and 1500~m, which is a heuristically good design, spanning between one and three times the depth of the interface. Since this starting design is closer to the final design, the SGD algorithm converges more quickly towards the final design. 

      The two SGD designs are compared to their respective initial designs and two benchmarks in Fig. \ref{fig:avo_onestep_comparison}. In both cases, but especially for the equispaced starting design, the \EIG (calculated using the $\text{NMC}_\text{re}$ method for better comparison) has increased substantially. Sequential construction ($\text{NMC}_\text{re}$ method using $200 * 2\tento{4}$ forward evaluations) is used here as a proxy for an optimal design. Since the value added by the 10th receiver is less than one per cent \citep{Guest2009-gi}, the \EIG of the sequential design will be very close to the global maximum. Both SGD designs perform slightly worse than the sequential construction design, which is most likely due to a combination of bias introduced by the variational approximation, choices in the learning rate, and small gradients due to the overdetermined nature of this design problem.

      The resulting designs can also be compared to an optimal design for the same setup but with a different model parameter prior pdf. Instead of a Gaussian, \citet{Guest2009-gi} used a Uniform distribution with upper and lower bound of 3000 and 4500~m/s respectively. Using the maximum entropy method, this results in a design with an \EIG of 2.95, which is better than the heuristic design but worse than the SGD and sequential construction designs, which shows the influence of the prior pdf in experimental design problems. Nevertheless, the design which outperforms the heuristic design even for a different prior pdf.
      
      The computational savings in this 10-dimensional design space are substantial. Even when repeated evaluations are saved, the SGD methods require around an order of magnitude fewer forward samples. When disregarding forward evaluations, the sequential design algorithm using the $\text{NMC}_\text{re}$ method requires $200 \times 10$ \EIG evaluations, each involving ${1\tento{4}}^2=1\tento{8}$ likelihood evaluations. In contrast, only $2-4\tento{5}$ forward and MDN evaluations are required to calculate the SGD experimental design. The evaluation of the variational family could be sped up substantially by using GPUs, which will be especially beneficial if more complex variational families are necessary.

    \subsection{\texorpdfstring{Interrogation for $CO_2$}{CO2}}

    \begin{table}
      \begin{tabular}{lllcl}
      Layer & Parameter  & Value  &  Unit  \\ \hline
      Upper Layer     & $\alpha_1$     & $2270 \pm 10$   & (m/s)      \\
                  & $\beta_1$      & $854 \pm 10$  & (m/s)      \\
                  & $\rho_1$       & $2100 \pm 10$   & (kg/m$^3$) \\
                    & d              & $1000 \pm 50$   & (m)        \\
      Lower Layer       & $K_{frame}$    & $2.56 \pm 0.77$ & (GPa)      \\
                  & $G_{frame}$    & $8.5 \pm 0.3$   & (GPa)      \\
                  & $\Phi$         & $0.37 \pm 0.02$ &            \\
                        & $K_{grain}$    & $39.3 \pm 1.4$ & (GPa)      \\
                  & $\rho_{grain}$ &  $44.8 \pm 0.8$ & (kg/m$^3$) \\
                        & $K_{brine}$    &  $2.31 \pm 0.07$ & (GPa)      \\
                  & $\rho_{brine}$ & $1030 \pm 20$   & (kg/m$^3$) \\
                       & $K_{co2}$      & $0.08 \pm 0.04$ & (GPa)      \\
                  & $\rho_{co2}$   &  $700 \pm 77$    & (kg/m$^3$)
      \end{tabular}
      \caption{
        Nuisance parameters in the AVO interrogation example. The quoted uncertainties correspond to the respective standard deviations.
        }\label{tab:avo_parameters}
      \end{table}

      As introduced in section \ref{subsec:srcloc_interrogation}, we will now demonstrate the use of variational methods for the design of experiments that answer a more practically interesting scientific question using AVO data. We focus on questions relating to the $CO_2$ saturation in a subsurface layer. The setup of the physical parameters represents a simplified model related to the Sleipner field \citep{Ghosh2020-xa, Dupuy2017-ln}. The upper layer is described by its seismic properties (P-wave velocity $\alpha_1$, S-wave velocity $\beta_1$), density $\rho_1$, and depth $d$, all with ranges given in Table \ref{tab:avo_parameters}. For the lower layer, the seismic properties are modelled using Gassmann fluid substitution \cite{Gassmann1951-ww, Smith2003-rc}, which can be used to calculate seismic parameters given properties of the drained frame (bulk modulus $K_\text{frame}$, shear modulus $G_\text{frame}$, porosity $\Phi$), mineral grains (bulk modulus $K_\text{grain}$, density $\rho_\text{grain}$), brine occupying the pore space (bulk modulus $K_\text{brine}$, density $\rho_\text{brine}$) and liquid $CO_2$ that replaces it (bulk modulus $K_\text{$CO_2$}$, density $\rho_\text{$CO_2$}$). Given all of these properties, only the saturation $S$ of the pore space by $CO_2$ is required in order to calculate the AVO effect, and is assumed to be uniformly distributed between 0 and 1. In contrast, all other parameters are assumed to be distributed according to a Gaussian with means and standard deviations given in Table \ref{tab:avo_parameters}.

      We must first calculate the pore fluid density and bulk modulus in order to apply the Gassmann equation. For this we use the Voigt (arithmetic) average 
      \begin{align}
        \rho_\text{fluid} &= S \rho_{CO_2} + (1-S) \rho_\text{brine} \\
        K_\text{fluid} &= S K_{CO_2} + (1-S) K_\text{brine}
      \end{align} 
      which gives a (stiff) upper bound on the bulk modulus. In real-world applications, a Reuss average or Voigt-Reuss-Hill average might be more suitable, but here the focus is on the optimal design algorithms rather than on details of the rock physical modelling. With the properties of the fluid at hand, the bulk modulus and density of the saturated rock can be modelled using the Gassmann equation
      \begin{align}
        K_\text{sat} &= K_\text{frame}+\frac{\left(1-\frac{K_\text{frame}}{K_\text{grain}}\right)^2}{\frac{\Phi}{K_\text{fluid}}+\frac{(1-\Phi)}{K_\text{grain}}-\frac{K_\text{frame}}{K_\text{grain}^2}} \\
        \rho_\text{sat} &= \Phi \rho_\text{grain} + (1 - \Phi) \rho_\text{fluid}
      \end{align}
      which can then be used to calculate the P-wave velocity of the lower layer. Using Gassmann's equation, we implicitly assume that the shear modulus of the lower layer is independent of the $CO_2$ saturation. The seismic properties of the lower layer can now be calculated as 
      \begin{align}
        \alpha_2 &= \sqrt{\frac{K_\text{sat} + \frac{4}{3} G_\text{frame}}{\rho_\text{sat}}} \\
        \beta_2 &= \sqrt{\frac{G_\text{frame}}{\rho_\text{sat}}}
      \end{align}
      which can then be used with the properties of the upper layer to calculate the P-wave reflection coefficient.

      \begin{figure}
        \centering
        \includegraphics[width=0.5\textwidth]{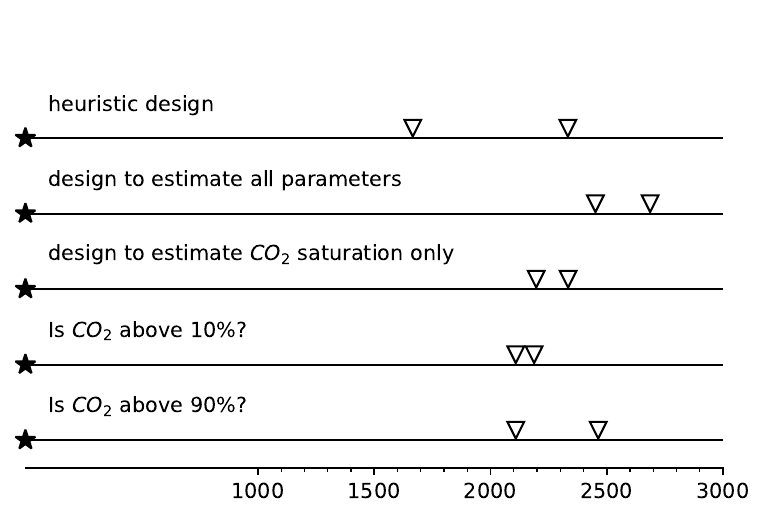}
        \caption{
          Offsets for two receiver designs for the AVO interrogation example. The heuristic design is chosen without optimisation the other four designs are optimal designs for different scientific questions.
        }\label{fig:avo_interrogation_comparison}
      \end{figure}

      We qualitatively compare optimised two station designs for different design aims and a heuristic design with two receivers equally spaced between one and three times the interface depth (1000m) at offsets of 1666 and 2333~m. In addition to the heuristic design, we calculated the design optimised to constrian all parameters given in Table \ref{tab:avo_parameters} and the $CO_2$ saturation using the $\text{NMC}_\text{re}$ method as a baseline for comparison. Sequential construction is used for this and all of the design optimisations in this subsection. To focus only on $CO_2$ saturation, we can see all parameters in Table \ref{tab:avo_parameters} as nuisance parameters. With the variational posterior method (with the same MDN as earlier in this section), we can now derive a design that is optimal for estimating the $CO_2$ saturation, with the resulting design shown in Fig. \ref{fig:avo_interrogation_comparison}. This design differs substantially from the heuristic design and the design constraining all model parameters. Since we have the mapping $T^{-1}$ (the Gassmann equation including nuisance parameters) available in this case, it would be possible to use extended (and computationally more expensive) versions of both the $\text{NMC}_\text{re}$ and variational marginal method for this interrogation design problem. This approach is not always possible if the scientific question is also a function of the $CO_2$ saturation.

      For some applications the exact value of the $CO_2$ saturation might not be of interest, but a key question is whether the saturation value is above or below a certain threshold. Changing the variational family of the variational posterior method to a neural network taking data as input, and predicting the probability of exceeding the threshold makes it possible to design experiments optimally suited for estimating whether the $CO_2$ saturation is above or below a certain threshold. The results of such an optimisation for a threshold of 0.1 and 0.9 are shown in Fig. \ref{fig:avo_interrogation_comparison}. The resulting designs are in a similar region as the one designed to constrain the value of the saturation, but deviate considerably to either focus on a more specific offset (threshold 0.1) or be spread further apart (threshold 0.9).

%% file: discussion.tex
Recent works have shown the need for approximations in experimental design methods for fully non-linear experimental design problems. Those approximations come in the form of linearised and Laplace methods which both assume that the forward model can be (locally) approximated by a linear model (\eg \citet{Long2015-wl, Krampe2021-sx, Wilkinson2006-kh, Maurer2017-tg, Carlon2020-os}), surrogates which approximate the forward model but put no constraints on model parameter prior or posterior pdf's (\eg \citet{Qiang2022-ow, Wu2022-oz, Huan2013-nf}), and functional approximations of the evidence, the posterior pdf or the mutual information between data and model parameters (\eg \citet{Coles2011-ks, Foster2019-rx, Kleinegesse2018-zv, Foster2019-rx, Kleinegesse2020-ht}). In this work we focused on variational methods which assume that that the evidence or posterior pdf can be described sufficiently well by a closed-form variational approximator. They do not require any modification to either the forward problem or the prior information on the model parameters, which makes them attractive for general-purpose applications.

While functional approximations introduce additional complexity compared to straightforward double-loop Monte Carlo estimators such as the NMC method, they have significant advantages, especially since the NMC method with reused samples, while working well in the presented examples, can perform suboptimal when compared to methods using functional approximations \citep{Englezou2022-ed}. Most importantly, they allow the design of experiments best suited to answer any scientific or applied question, provided a mapping from model space to the relevant target space can be defined. While this is important in its own right, the typical low dimensionality of the target space could allow experimental design methods to scale to substantially larger problems than currently possible.

Another significant advantage is the straightforward application of SGD design optimisation using \EIG lower bounds (variational posterior and InfoNCE method in this study) if gradients with respect to the design parameters are available. While this is also possible using NMC or Laplace methods, they need a significant number of inner loop samples \citep{Goda2020-uo} or require an estimation of the maximum a posteriori estimate \citep{Carlon2020-os}, respectively, at each gradient descent step. The Laplace method can be seen as a special case of the variational posterior method in which the variational family is replaced by a multivariate Gaussian derived using the Hessian matrix of the linearised forward problem. Therefore they should introduce a similar bias as incurred when using a well-trained MDN predicting one Gaussian with full covariance matrix. The same goes for the consistent extensions of both methods, the VNMC method of \citet{Foster2019-rx} and the Laplace-based importance sampling estimator of \citet{Carlon2020-os, Englezou2022-ed}. Unlike the variational methods, the Laplace-based methods are inherently restricted to a Gaussian posterior pdf.

The $D_N$ method performs well in the benchmarks presented in this paper. It shows the advantages of using a variational family which can be fit easily and whose information content can be evaluated analytically. Extending the method to be applicable for (some) interrogation design problems could benefit large-scale geophysical applications. It would provide a cheap and robust method that can be applied to many applications. If an inverse mapping of the target function exists or can be approximated, the variational marginal-likelihood method of \citet{Foster2019-rx} can be used to extend the $D_N$ method to a subclass of interrogation problems. However, as has been demonstrated, the $D_N$ method can lead to non-optimal designs if the assumption of Gaussian evidence is violated substantially. For complex high-dimensional problems, deciding whether a problem is ill-suited for the $D_N$ method will be difficult, however it is clear, that special care should be taken if the prior is multimodal or if the forward function is known to show strong non-linearity. Nevertheless, apart from in extreme cases, there is reason to believe that $D_N$ design will provide at least above-average (compared to randomly selected) designs, at substantially reduced cost compared to most other design methods, and it is the only method that can produce robust designs using of the order of 10 to 100 forward evaluations. 

Specifying variational mappings is a non-trivial task, and care needs to be taken to provide a sufficiently expressive yet computationally tractable one. Some of those problems can be alleviated using lower bounds on the \EIG solely parametrised by neural networks \citep{Kleinegesse2021-aq, Guo2021-hx}. Those have the same advantages as the variational posterior method but are easier to specify and can be constructed to be consistent (converge to the true \EIG given enough samples) and with low variance \citep{Guo2021-hx}, making them well suited for SGD design optimisation. They are promising candidates for large-scale interrogation experimental design problems.

While some results shown here could have been derived qualitatively using physical intuition, calculating the exact design requires a quantitative framework, as discussed in this paper. This is especially true for complex 3D scenarios involving many receivers, complex priors, noise distributions and physical models, where heuristics and intuition break down.

%% file: conclusions.tex
In this paper, we have introduced variational methods to Geophysics, we have discussed their potential benefits and challenges, and placed them into context amongst linearised and other more established methods. We also briefly introduced the use of mutual information lower bounds for experimental design. The examples were chosen to illustrate the main concepts and encourage the use of these methods in geophysical applications.

We have compared different methods for estimating the value of an experiment and show that the naive neste Monte Carlo (NMC) method is impractical for even small-scale geophysical problems due to the large number of samples required. In contrast, the variational methods and the NMC method with reused inner loop samples perform similarly well for both the seismic source location and amplitude-versus-offset (AVO) design problems. All three methods can fully account for the effects of nonlinearity in the physical process, but which method to is preferred depends on the problem at hand. In line with previous studies, we have also shown that the $D_N$ method can substantially increase \EIG of an experiment while being exceptionally computationally cheap and straightforward to apply. Using the framework of variational methods, we also demonstrated how this method could be extended to account for any noise distribution. However, we have also shown that the $D_N$ method can lead to non-optimal results in the presence of multimodality in the evidence, and that to determine when it will fail is difficult without also applying other, more expensive design methods.

We have also demonstrated how lower bounds on the expected information gain can be used to design interrogation experiments, that provide the best possible answer to any specific questions of interest. We argue that this focused approach is more efficient and uses resources better than the traditional optimal design if a specific research question is of scientific interest. We also show that the optimal design can change substantially depending on the question posed. 

Using AVO analysis as an example of a highly non-linear geophysical process, we demonstrate the applicability and computational saving enabled by deploying stochastic gradient design optimisation. This is especially relevant for high-dimensional designs that collect a large number of data. Even for expensive forward problems a small number of gradient descent steps can be used to refine heuristic designs to a specific problem at hand.

All methods used have been implemented in a Python package\footnote[1]{Available under \url{https://github.com/dominik-strutz/GeoBOED}} to enable the use of OED for other researchers. Currently, it is in an early stage and lacks documentation. Still, it will be updated consistently over the coming years, and user-friendly documentation and tutorials will be added.

%% file: data_availability.tex
This study only uses synthetic data. The code to generate the figures in section \ref{sec:applications} is available here: \url{https://github.com/dominik-strutz/VarBEDfGP}

%% file: acknowledgements.tex
The implementations of the OED algorithms in this work would not have been possible without extensive use of open-source software. Not all of them have been included in the respective sections to ease readability. All the code was written in Python \citep{Van_Rossum2011-ep}, the libraries PyTorch \citep{Paszke2019-lv} and Zuko \citep{Rozet2023-zy} were used to process probability distributions and implement the variational families, NumPy \citep{Harris2020-nt} was used for general data processing and Matplotlib \citep{Hunter2007-jw} for plotting.

This project has received funding from the European Union’s Horizon 2020 research and innovation programme under the Marie Skłodowska-Curie grant agreement No 955515 – SPIN ITN (www.spin-itn.eu)